\documentstyle[emulateapj]{article}

\slugcomment{The Astrophysical Journal, in press}

\begin{document}

\title{Diffuse Gas and LMXBs in the {\it Chandra} Observation of the S0 Galaxy NGC~1553}

\author{Elizabeth L. Blanton\altaffilmark{1},
Craig L. Sarazin\altaffilmark{1},
and
Jimmy A. Irwin\altaffilmark{2,3}}

\altaffiltext{1}{Department of Astronomy, University of Virginia,
P. O. Box 3818, Charlottesville, VA 22903-0818;
eblanton@virginia.edu, sarazin@virginia.edu}

\altaffiltext{2}{Department of Astronomy, University of Michigan,
Ann Arbor, MI 48109-1090; jirwin@astro.lsa.umich.edu}

\altaffiltext{3}{$Chandra$ Fellow}

\begin{abstract}
We have spatially and spectrally resolved the sources of X-ray emission
from the X-ray faint S0 galaxy NGC~1553 using an observation from the
$Chandra$ X-ray Observatory.
The majority (70\%) of the emission in the $0.3 - 10.0$ keV band is diffuse,
and the remaining 30\% is resolved into 49 discrete sources.
Most of the discrete sources associated with the galaxy appear to be
low mass X-ray binaries (LMXBs).
The luminosity function of the LMXB sources is well-fit by a broken
power-law with a break luminosity comparable to the Eddington luminosity for 
a 1.4$M_{\odot}$ neutron star.
It is likely that those sources with luminosities above the
break are accreting black holes and those below are mostly neutron
stars in binary systems.
Spectra were extracted for the total emission, diffuse emission, and
sum of the resolved sources;
the spectral fits for all require a model including both a soft and
hard component.
The diffuse emission is predominately soft while the emission from the
sources is mostly hard.
Approximately 24\% of the diffuse emission arises from unresolved
LMXBs, with the remainder resulting from thermal emission from hot
gas.
There is a very bright source at the projected position of the nucleus
of the galaxy.
The spectrum and luminosity derived from this central source are consistent
with it being an AGN;
the galaxy also is a weak radio source.
Finally, the diffuse emission exhibits significant
substructure with an intriguing spiral feature passing through the
center of the galaxy.
The X-ray spectrum and surface brightness of the spiral feature are
consistent with adiabatic or shock compression of ambient gas, but not with
cooling.
This feature may be due to compression of the hot interstellar gas by
radio lobes or jets associated with the AGN.
\end{abstract}


\section{Introduction} \label{sec:intro}

There is evidence from their X-ray spectra that a variety of different
emission mechanisms contribute to the X-ray luminosity of early-type
galaxies.
Elliptical and S0 galaxies exhibit varying X-ray spectra that depend on 
the ratio of the galaxy's
X-ray-to-optical luminosity ($L_{X}/L_{B}$), (Kim, Fabbiano, \& Trinchieri
1992).  X-ray
bright E and S0 galaxies (those with high $L_{X}/L_{B}$) emit thermally 
with a spectrum that is well modeled
by an optically thin plasma with a temperature of $kT \approx 1$ keV (Forman,
Jones, \& Tucker 1985; Canizares, Fabbiano, \& Trinchieri 1987).  X-ray
faint E and S0's (those with low $L_{X}/L_{B}$) 
have spectra that require a model comprised of two components,
soft emission with $kT \approx 0.2$
(Fabbiano, Kim, \& Trinchieri 1994)
and a hard component
(Matsumoto et al.\ 1997).
The hard component has been fit either as thermal bremsstrahlung with
$kT \ga 5 $ keV (Matsumoto et al.\ 1997)
or a power-law
(Allen, di Matteo, \& Fabian 2000).
The hard component of the spectra for X-ray faint E and S0
galaxies is roughly proportional to the optical luminosity of the galaxy,
suggesting that its origin is low-mass X-ray binaries (LMXBs) such as those
seen in our Galaxy (Trinchieri \& Fabbiano 1985).
In some galaxies, a central active galactic nucleus (AGN) may be an important
source of hard X-ray emission
(Allen et al.\ 2000).
The origin of the soft component remains uncertain (Fabbiano, Kim, \& Trinchieri 1994;
Pellegrini 1994; Kim et al.\ 1996; Irwin \& Sarazin 1998a,b).
It might be due to interstellar gas, LMXBs, some other stellar component,
or some other source.
Irwin \& Sarazin (1998a,b) argued that LMXBs produce much of the soft emission
in the most X-ray faint galaxies, while the contribution of interstellar gas
increases as $L_{X}/L_{B}$ increases.
In X-ray faint E and S0 galaxies, much of this hot 
gas may have been lost in galactic winds or as a result of ram pressure 
stripping by intragroup or intracluster gas.

Until now, X-ray detectors have not had sufficient spatial resolution to 
resolve the LMXBs from diffuse emission in most external galaxies.
The {\it Chandra} X-ray Observatory has given us the opportunity to resolve
these components spatially and spectrally and to uncover the source of
emission from X-ray faint E and S0 galaxies.
In this paper, we present a {\it Chandra}
observation of the S0 galaxy NGC~1553.  Its $L_{X}/L_{B}$ ratio puts it on the
bright end of the class of X-ray faint ellipticals and S0's 
(Irwin \& Sarazin 1998b).  Although other galaxies have lower $L_{X}/L_{B}$
ratios, NGC~1553 was chosen because its high X-ray flux and proximity make it
a good candidate for resolving LMXBs with $Chandra$.
Another X-ray faint galaxy, NGC~4697, with a lower $L_{X}/L_{B}$ ratio was
also observed (Sarazin, Irwin, \& Bregman 2000).
NGC~4697 is closer with a lower X-ray flux and will sample
the faint end of the LMXB luminosity function.  
The $L_{X}/L_{B}$ ratio and X-ray colors of NGC~1553 put it in a class of 
galaxies that 
Irwin \& Sarazin (1998b) hypothesized may have some contribution of hot ISM to
the spectrum, that could be fit by a thermal model with $kT = 0.3 - 0.6$ and
metallicity of 10-20\% of solar along with a stellar component.

NGC~1553 is interacting with a companion, NGC~1549, as evidenced by shells of
emission seen in the optical.  The two galaxies are separated by 11$\farcm$8
(Bridges \& Hanes 1990).  NGC~1553 is known to have cool gas and dust 
associated with it.  It was detected in the infrared in all four $IRAS$ bands
studied in Bally \& Thronson (1989).  In addition, it is a weak radio
source with a flux density of 10 mJy at 843 MHz (Harnett 1987).  A dust lane
was observed at small scales ($\approx 3^{\prime\prime}$ across) at the 
center of the galaxy with HST
(Quillen, Bower, \& Stritzinger 2000).
Previous X-ray observations with
the $ROSAT$ HRI
(Trinchieri, Noris, \& di Serego Alighieri, 1997)
revealed substructure in the X-ray
map, but the resolution was not high enough to distinguish individual LMXBs.

The distance to NGC~1553 is somewhat poorly known, and reported values
vary widely.
However, NGC~1553 is known to be interacting with (and
therefore at the same distance as) the elliptical galaxy NGC~1549.
Consistent measurements for the distance to NGC~1549 of 24.2 Mpc are given by
Faber et al.\ (1989; assuming $H_{\circ} = 50$ km s$^{-1}$ Mpc$^{-1}$)
and Tonry (2000) using 
the $D_{n}-\sigma$ and surface brightness fluctuations methods, respectively. 
We thus adopt 24.2 Mpc as the distance to NGC~1553.

\section{Observation and Data Reduction} \label{sec:data}

NGC~1553 was observed with {\it Chandra} on 2000 January 2-3 for a total of
33,882 s.
The center of the galaxy was positioned on the ACIS-S3 CCD, with
an offset of one arcmin from the nominal pointing for the S3.
The S3 was selected for its sensitivity at low energies.
In addition to the S3, data were received from the
ACIS I2, I3, S2, and S4 CCDs.
In the analysis that follows, we use data from the S3 chip only.
The events were telemetered in Very Faint mode, the data were collected
with frame times of 3.2 seconds, and the CCD temperature was -110 C.
Only events with $ASCA$ grades of 0,2,3,4, and 6 were included.
Bad pixels, bad columns, and columns to either side of bad 
columns and node boundaries were excluded.
Event PI values and photon energies were determined using the
acisD1999-09-16gainN0004.fits gain file.

\begin{figure}[t!]
\plotone{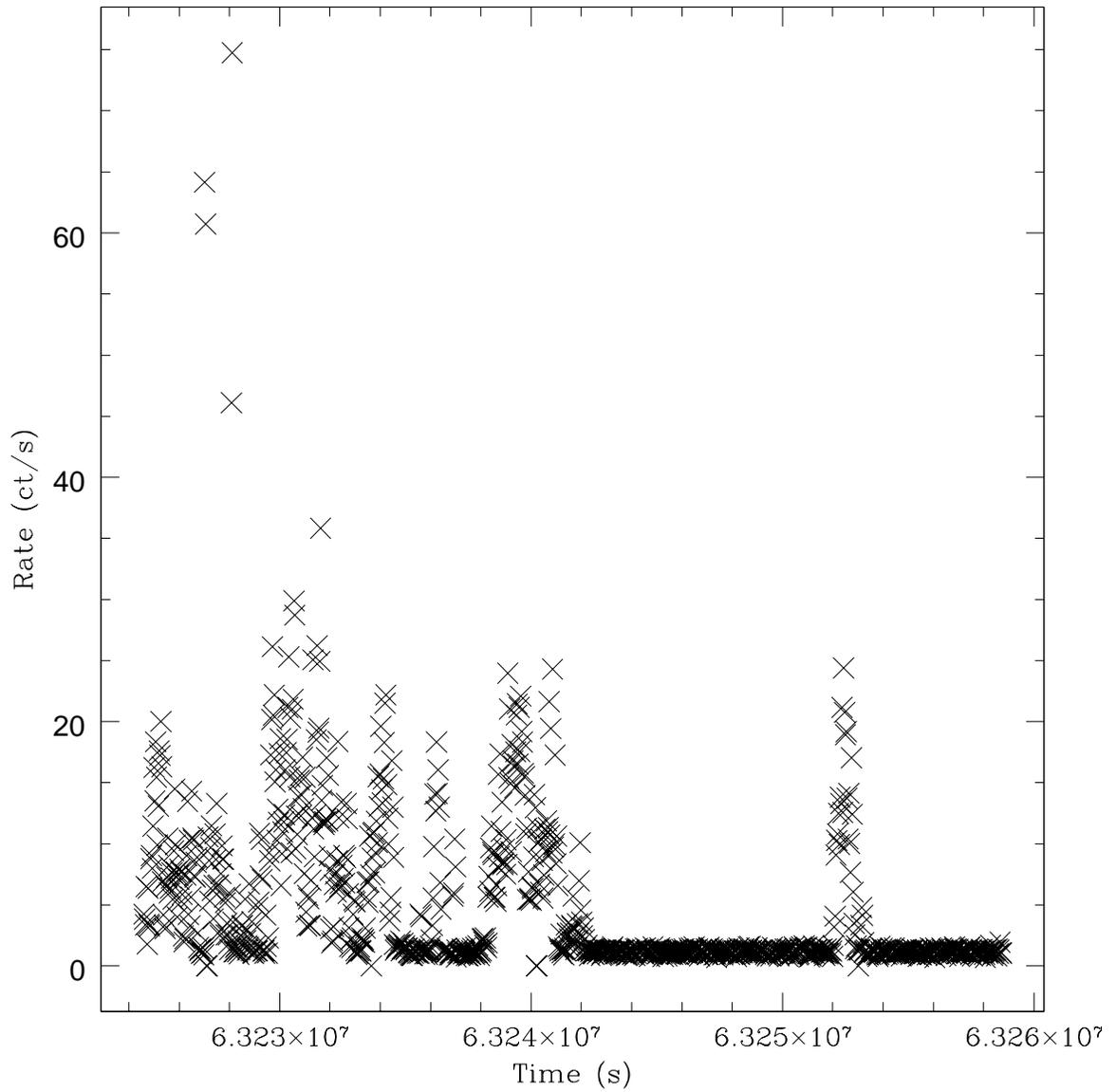}
\caption{Light curve of the total X-ray emission detected by the ACIS S3
chip in the 0.3-10.0 keV band.
The time is measured in seconds.
Background flares are evident, particularly during the 
first half of the exposure.
The quiescent rate is about 1.19 ct s$^{-1}$.
\label{fig:lightcurve}}
\end{figure}

{\it Chandra} is known to experience periods of high background
(``background flares''), which are particularly significant for the
S3 chip (Markevitch 2000; Markevitch et al.\ 2000).
We searched our data for such flares and found that, 
unfortunately, much of the exposure was affected by them.
A light curve for the entire S3 chip for events with energies in the
range 0.3 - 10.0 keV is displayed in Figure~\ref{fig:lightcurve}.
There are 1000 time bins of 34.17 s.
The average quiescent source plus background rate is about
1.19 ct s$^{-1}$, of which 0.85 ct s$^{-1}$ are background
(Markevitch 2000)
and 0.34 ct s$^{-1}$ are due to NGC~1553.
The background flares are most prevalent during the
first half of the exposure, where the background rate reached upwards of 60
ct s$^{-1}$.
Ideally, one would like to remove all of the exposure where the background
is increased significantly above the quiescent background
(Markevitch 2000).
Unfortunately, this would remove most of the exposure.
In order to retain as much of the exposure as possible, while 
also keeping the signal-to-noise ratio as high as possible, we excluded data 
from our analysis that had a total S3 count rate greater than 5 ct s$^{-1}$.  
With this cutoff value, we eliminated 81\% of the background and only 31\% of
the exposure.
The resulting total exposure, corrected for background flares, is 23,192 s.  
We also searched for periods of incorrect aspect solution, and none were found.

In our analysis of the diffuse emission in NGC~1553, we determine the
background from a circular region as far away as possible from the 
galaxy on the S3 CCD.
Unfortunately, the elevated background in our observations, even after
removing the strongest background flares, meant that we cannot use
the available blank sky background fields, which were all taken during
times of quiescent background only
(Markevitch 2000).
In addition, the other backside illuminated CCD (the S1) was turned off
during this observation due to the FEP0 problem, so we could not take
background from the S1.
It is possible that even the outermost regions of the S3 chip may still contain
some emission from NGC~1553, so our analysis may underestimate the total amount
of diffuse emission.
However, examination of the $ROSAT$ PSPC image of 
NGC~1553 reveals that there does not appear to be galaxy emission in our
background region.
The analysis of resolved sources was not affected by this, as
the background was determined locally so that both true background and
superposed diffuse emission could be subtracted.

This observation was processed at a time when the standard pipeline
processing introduced a boresight error of about 8\arcsec\ in the
absolute positions of X-ray sources.
We made an approximate correction for this error.
The resulting position for the bright X-ray source near the center
of NGC~1553 agreed with the {\it Hubble} Space Telescope (HST)
position (data kindly provided by A. Kundu) of the center of the galaxy to
1\farcs9.
There are a number of arguments which indicate that the central X-ray
source is associated with the active nucleus of the galaxy
(\S~\ref{sec:agn}).
Thus, we corrected the X-ray positions so that the central X-ray
source coincided with the nucleus of NGC~1553 as determined with HST.
We also found a very good agreement (0\farcs50) between the positions of
three X-ray sources and three globular clusters in NGC~1553 found
with HST
(A. Kundu, private communication).
Before correction, all three globulars were displaced from their X-ray
source in the same direction as the optical and X-ray position of the
central AGN.

The initial data reduction was done with the {\it Chandra} analysis package
CIAO.
Some of the image analysis was done with IRAF/PROS.
The spectra were extracted using software very kindly provided by
Alexey Vikhlinin
(Vikhlinin, Markevitch, \& Murray 2000).
The spectra were analyzed using XSPEC.

\subsection{X-ray Image} \label{sec:image}

\begin{figure}[t!]
\plotone{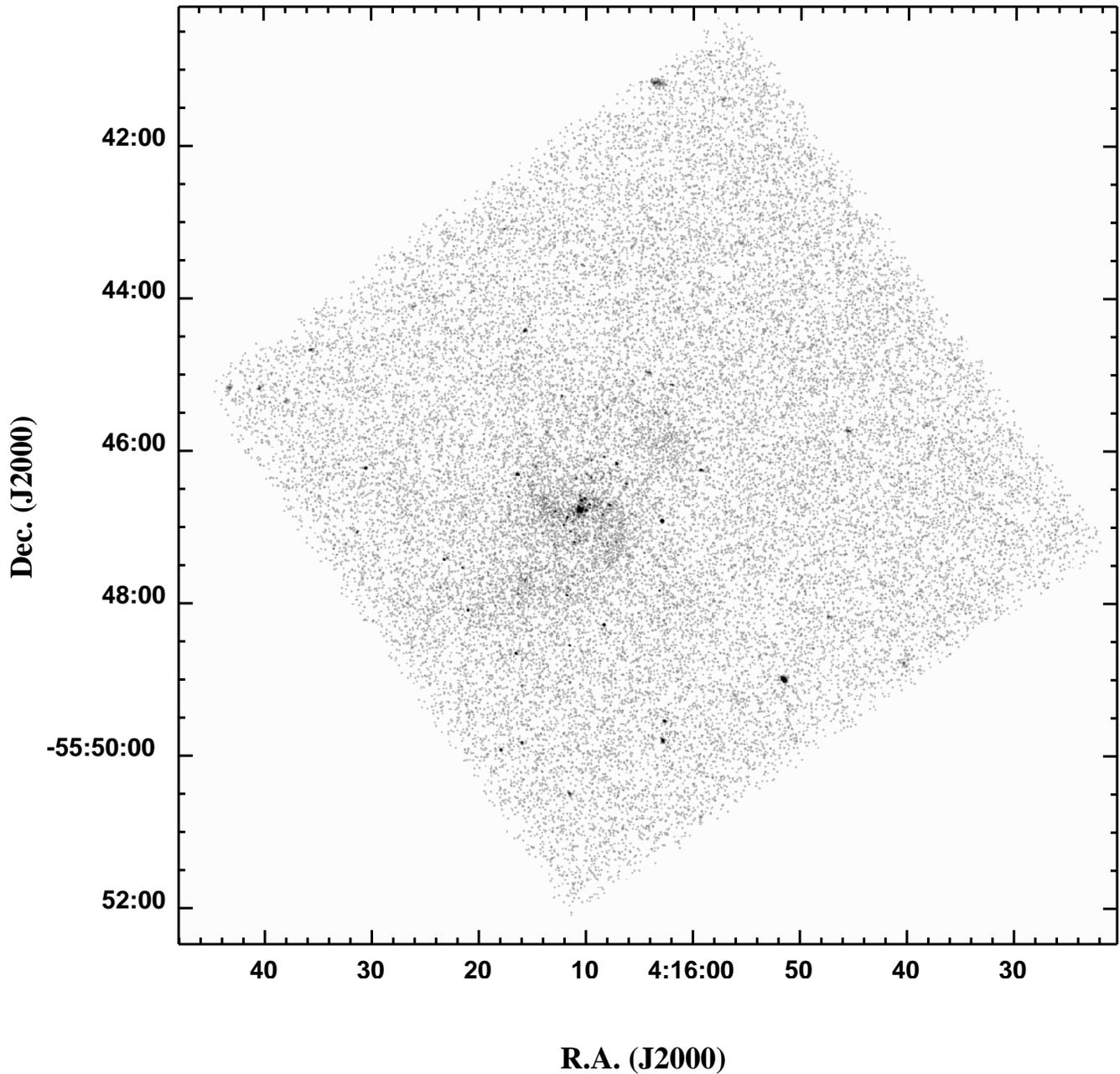}
\caption{Slightly smoothed ($\sigma = 0\farcs492 = 1$ pixel Gaussian)
{\it Chandra} S3 image of NGC~1553. 
Discrete sources as well as asymmetric diffuse emission are apparent.
The center of the galaxy is located to the SE of the detector center,
and is coincident with the brightest point source in the field.
\label{fig:xwhole}}
\end{figure}

Figure~\ref{fig:xwhole}
shows the {\it Chandra} S3 image of NGC~1553 in the 0.3 - 10.0 keV energy
band.  The image has not been corrected for exposure or background.  It has
been smoothed slightly with a
$\sigma = 0\farcs492$ (1 pixel) Gaussian so that features may be seen
more easily.  Discrete sources are evident and are concentrated at the center
of the galaxy.
The brightest source in the field is located at the center of NGC~1553,
and is probably due to the active nucleus of the galaxy.

In addition, diffuse emission is apparent in Figure~\ref{fig:xwhole}.
In general, the diffuse emission is elongated at a position angle of
$\sim140^\circ$, which is similar to that of the optical light
(Fig.~\ref{fig:optical} below).
The diffuse component is seen more
clearly in Figure~\ref{fig:xsmooth}
which displays the adaptively smoothed 
$4^{\prime}\times4^{\prime}$ central region of the galaxy.  The smoothed image
has a minimum signal-to-noise ratio (SNR) of three per smoothing beam and was
corrected for background and exposure.  The grayscale is logarithmic and ranges
from $3.8\times10^{-7}$ to $3.0\times10^{-3}$ ct pix$^{-1}$ s$^{-1}$.

The diffuse X-ray emission is 
asymmetric
with extensions to the NW and SE.
In addition there is a bright ``spiral'' feature in the diffuse emission
that runs SW to NE through the center of the galaxy.
This spiral is approximately perpendicular to the optical axis.  

\section{Resolved Sources} \label{sec:sources}

Discrete sources were found on the ACIS S3 image by running the CIAO 
$wavdetect$
wavelet detection algorithm.
The source detection threshold was set at $10^{-6}$,
implying that $\la 1$ false source (due to a statistical fluctuation)
would be detected within the area of the S3 image.
The detection limit for the sources corresponds approximately
to a count rate of $4.3 \times 10^{-4}$ ct s$^{-1}$.
The sources were checked on the X-ray image,
and the $wavdetect$ count rates were
verified with aperture photometry in IRAF/PROS.  The count rates from the
two methods agreed well in all cases.
A total of 49 sources were found that exceeded the detection threshold;
they are listed in
Table~\ref{tab:sources}.
The columns give the source number, its name, its position, its projected distance
$d$ from the center of NGC~1553, the count rate and error, the signal-to-noise
ratio $SNR$ of the detection, the X-ray luminosity $L_X$ assuming the sources
are located at the distance of NGC~1553 (\S~\ref{sec:lumfunc} below),
and notes.
The sources are listed in order of increasing distance from the center of
NGC~1553.
Based on background source number counts from the 
deep {\it Chandra} observations of Mushotzky et al.\ (2000) and
Brandt et al.\ (2000), we expect that 6 - 12 of the sources in NGC~1553 may be
unrelated to the galaxy.

We examined the positions of the resolved X-ray sources on the Digital Sky
Survey (DSS) image of the field of NGC~1553.
Figure~\ref{fig:optical} shows an overlay of the X-ray source
positions onto the DSS image.
We found six possible optical identifications from the DSS,
which are noted with ``DSS'' in column 9 of Table~\ref{tab:sources}.
The DSS positions are consistently offset to the SW from the {\it Chandra}
positions, with an average offset of 2$^{\prime\prime}$.
These associations are mainly in the outer parts of the field, as it is
difficult to detect faint optical sources in the inner regions of
NGC~1553 using the DSS.

We also compared the positions of our X-ray sources with a catalog
of globular cluster candidates in NGC~1553 based on HST observations
(A. Kundu, private communication).
The three {\it Chandra} sources that have positions coincident
(to within 0\farcs5) with those of globular clusters are Srcs.~4, 11, and 23
(CXOU J041609.6-554642, CXOU J041611.0-554712, and CXOU J041621.5-554731).
These sources are indicated by ``HST'' in the notes column.
Based on the numbers of globular clusters and X-ray
sources found within the region covered by the HST observation, we expect only
0.05 chance matches within 0\farcs5.
We therefore conclude that all three of these associations are probably real.

Comparison with the 843 MHz radio map in Harnett (1987) reveals extended radio
emission with a flux density of 10 mJy coincident with the center of NGC~1553.
In addition, there is a bright nearby point source with a flux density of
166 mJy at a distance of approximately two arcminutes from the galaxy center.
The position of this radio sources coincides (within 1 arcsec) of the
$Chandra$ detected Src.~27, as well as an optical object on the DSS.  It is
likely a background object.

The second brightest resolved source detected on the S3 chip is Src.~38 
which is more than an order of 
magnitude brighter than most of the other detected sources, and lies 
$3\farcm5$ to the SW of the center of the galaxy.  It is coincident with an
optical object on the DSS.  This source was
also seen in $BeppoSAX$ data by Trinchieri et al.\ (2000), and over the course
of their two observations, it varied in brightness.  Given its
distance from the center of the galaxy and its brightness, it is likely a
background object.
Trinchieri et al.\ (2000) also marginally detected a hard X-ray
source ($\sim$10--100 keV) with the PDS in the field of NGC~1553.
The lack of spatial resolution of the PDS means
that a counterpart could be anywhere within approximately 75 arcmin,
which is much larger than the field of view of our $Chandra$ observation.
None of the $Chandra$ sources has a hard X-ray flux (say, 5--10 keV)
which is within an order of magnitude of the PDS source, so it is
unlikely that this very hard X-ray emission is produced by any of the
sources we detected.

\subsection{Hardness Ratios} \label{sec:hardness}

\begin{figure}[t!]
\plotone{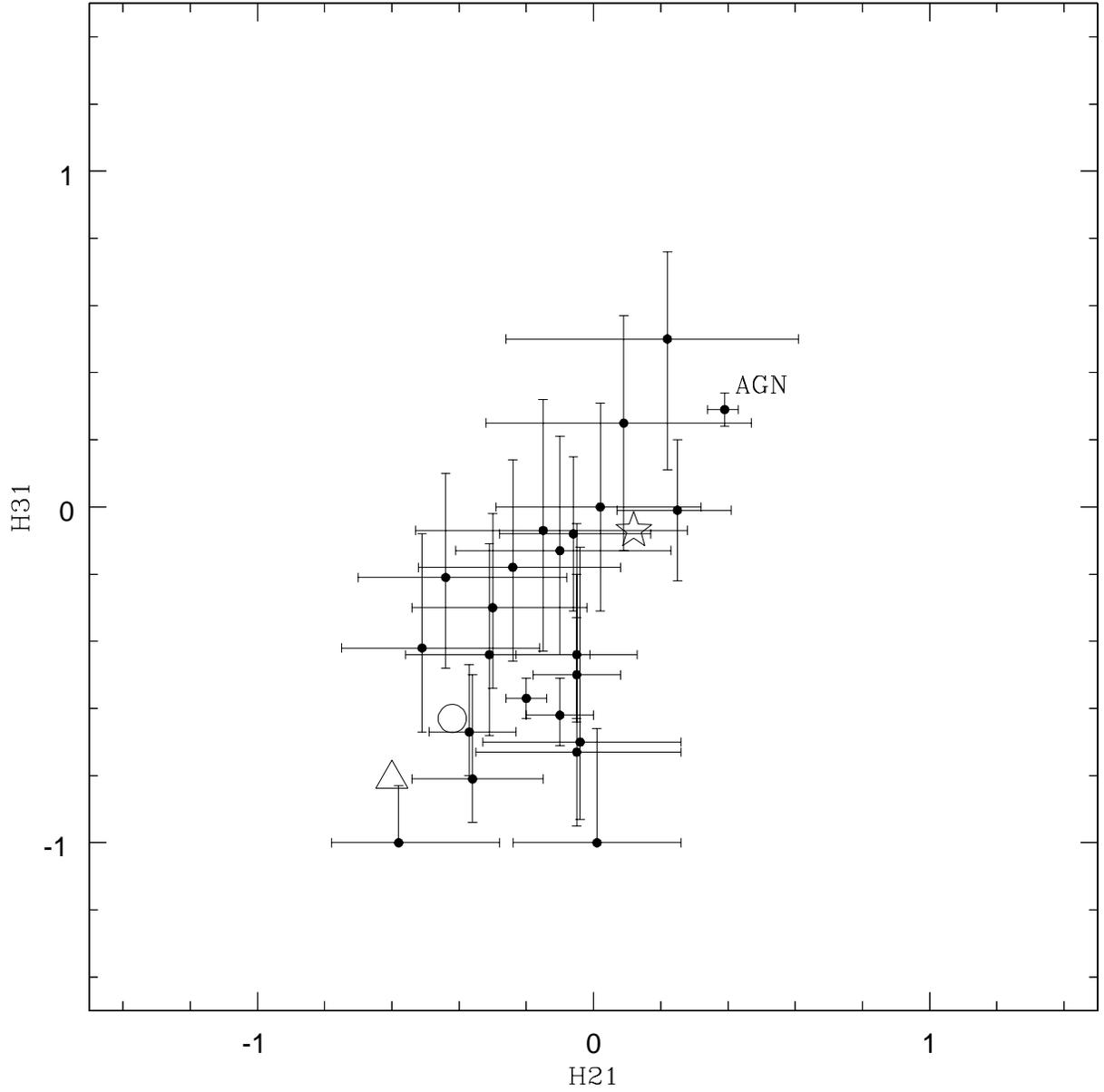}
\caption{X-ray hardness ratios of the discrete sources detected with more
than 20 counts in the S3 CCD in the 0.3 - 10.0 keV band.
The hardness ratios are $ H21 = (M-S)/(M+S)$, and
$H31 = (H-S)/(H+S)$.
The point marked ``AGN'' is the central source in NGC~1553.  For comparison,
the open circle shows the hardness ratios for the total emission, the open
triangle shows those for the diffuse emission, and the open star shows the
ratios for the sum of the discrete sources.
\label{fig:colors}}
\end{figure}

We first study the crude spectral properties of the resolved sources
using hardness ratios, which have the advantage that they can be determined
for fainter sources than detailed spectra.
For consistency with the previous {\it Chandra} study of NGC~4697
(Sarazin, Irwin, \& Bregman 2000), we use
the three energy bands: soft S 0.3--1.0 keV,
medium M 1.0--2.0 keV, and hard H 2.0--10.0 keV.
As in Sarazin, Irwin, \& Bregman (2000),
we define two X-ray hardness ratios $ H21 = (M-S)/(M+S)$, and
$H31 = (H-S)/(H+S)$.
For those resolved sources that have more than 20 counts, the hardness ratios
and their errors are shown in Figure~\ref{fig:colors}.
For comparison, the
hardnesses of the total emission (diffuse plus sources) are
$H21=-0.42$ and 
$H31=-0.63$.
Isolating the diffuse emission gives hardness ratios of $H21=-0.60$ and
$H31=-0.81$, while the sum of the resolved sources yields $H21=0.12$
and $H31=-0.07$.
Most of the resolved X-ray sources have colors which are similar to
the sources detected with {\it Chandra} in the X-ray faint elliptical
NGC~4697
(Sarazin, Irwin, \& Bregman 2000).
Unlike NGC~4697, no supersoft sources were found.

The central source (Src.~1, CXOU J041610.5-554646), which is the
brightest source in the field and whose position agrees with the
nucleus of the galaxy, has hard X-ray colors,
$H21=0.39$ and $H31=0.29$.
This suggests that it is a strongly absorbed AGN rather than an X-ray
binary in the galaxy (\S~\ref{sec:agn}).

\subsection{Variability} \label{sec:vary}

We searched for temporal variations in the X-ray emission from the individual
resolved sources in the 0.3--10.0 keV energy band.
We are limited in such searches by the relatively small number of photons
detected from most of the sources.
The timing resolution is also limited by the frame time of 3.2 s.
Thus, we only performed a relatively simple test;
we used the Kolmogoroff-Smirov (KS) statistic to test the null hypothesis
that the source plus 
background rate is uniform over the active exposure time.
We consider sources to be possibly variable only if a uniform emission
rate can be ruled out at the 90\% confidence level.
Using this technique, we find that five of the resolved sources may be 
variable.
They are Srcs.~4, 17, 36, 40, \& 41
(CXOU J041609.6-554642,
CXOU J041602.9-554655,
CXOU J041617.9-554955,
CXOU J041545.5-554543,
\&
CXOU J041611.5-555029), which
have probabilities of 7.5\%, 2.2\%, 2.4\%, 5.2\%, and 4.6\%, respectively, 
that they are not variable.
The first of these sources is also 
coincident with a globular cluster candidate detected with HST.
However, since we tested 49 sources, we expect 10\% or $\sim$5 to appear to
be variable due to random fluctuations.

\section{Luminosities and Luminosity Function of Resolved Sources}
\label{sec:lumfunc}

The count rates for the sources were converted into X-ray luminosities
$L_X$ in the 0.3--10.0 keV energy band assuming all of the sources are at the
distance of NGC~1553 of 24.2 Mpc.
To determine the conversion from count rate to unabsorbed energy flux in
the band 0.3--10 keV, we used the best fitting spectral model for
the cumulative spectrum of all of the resolved sources (excluding the central
source and Src.~38 which is very bright and likely a background object).  
We fit the 
spectrum in the energy range 0.7--10.0 keV, where we are confident that the
response is accurate, with a model including absorption fixed at the 
Galactic value, a soft blackbody component and a hard power-law component
(\S~\ref{sec:spec_sources}).  We then took the unabsorbed flux from
0.3--10.0 keV without refitting the spectrum for the expanded energy
range.
The resulting conversion between count rate and luminosity is
$4.91\times10^{41}$ ergs ct$^{-1}$.
The luminosity values are given in column 8 of Table~\ref{tab:sources}.
The source luminosities range
between $1.9\times10^{38}$ and $1.8\times10^{40}$ erg s$^{-1}$.

\begin{figure}[t!]
\vskip 5.0truein
\includegraphics{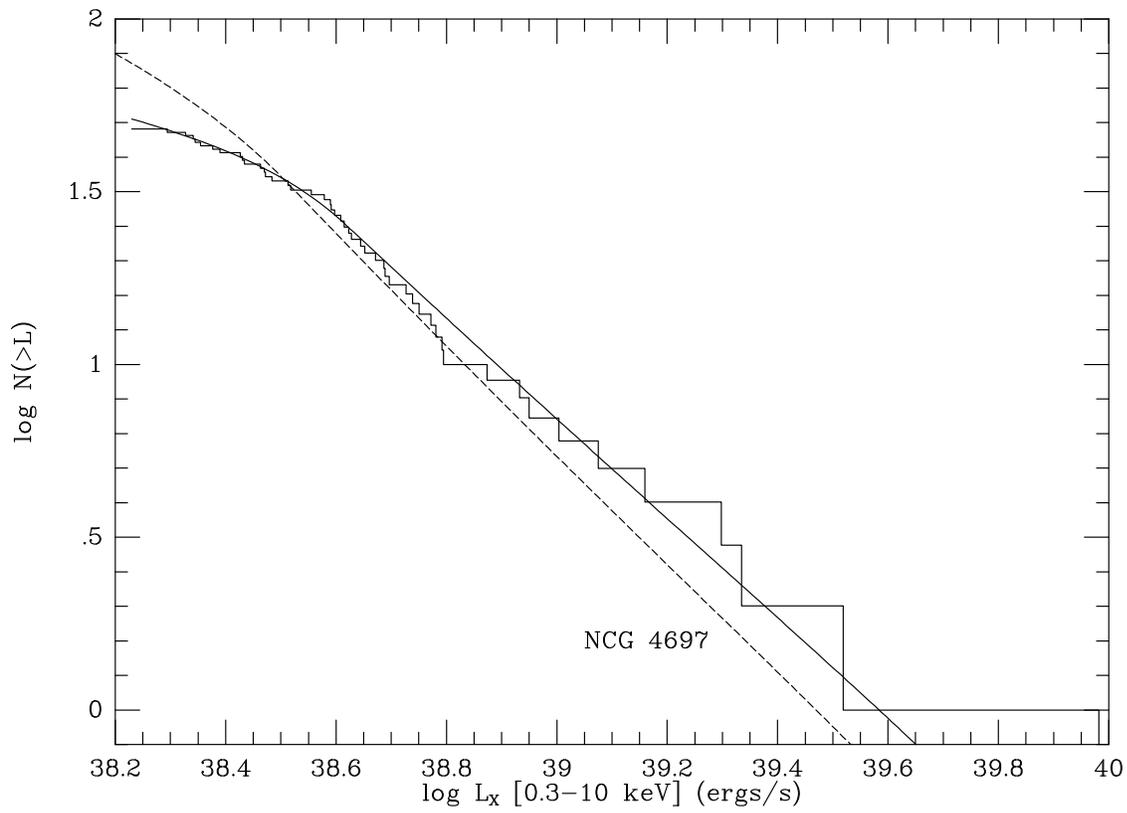}
\caption{The histogram is the luminosity function of the discrete X-ray sources
in NGC1553.
The solid curve represents the best-fitting broken power-law function.
For comparison, the dashed line shows the broken power-law fit for NGC~4697
(Sarazin et al. 2000).
\label{fig:lumfunc}}
\end{figure}

A luminosity function was constructed for the sources and is plotted as a
histogram in
Figure~\ref{fig:lumfunc}.
The central source (Src.~1) was not included in this luminosity function.
We fit the luminosity function of the resolved sources to either
a power-law or broken power-law model for the galactic sources,
plus the expected contribution of background and foreground sources,
determined from the deep blank field {\it Chandra} observations of
Brandt et al.\ (2000)
and
Mushotzky et al.\ (2000).
The background sources are assumed to follow
the number versus flux relation derived from {\it ROSAT} deep fields
(Hasinger et al.\ 1998).
A single power-law did not produce an acceptable fit to the observed
luminosity function;
it could be rejected at the $>$94\% confidence level.
Fitting a broken power-law, however, 
resulted in a good fit, which is shown as the continuous curve
in Figure~\ref{fig:lumfunc}.
The broken power-law luminosity function takes the form:
\begin{eqnarray}
\frac{dN}{dL_{38}} = N_{\circ}\left(\frac{L_{38}}{L_{b}}\right)^{-\alpha} ,
\end{eqnarray}
where $\alpha=\alpha_{l}$ for $L_{38} \le L_{b}$, $\alpha = \alpha_{h}$
for $L_{38} > L_{b}$, and $L_{38}$ is the 0.3--10 keV X-ray luminosity in units
of 10$^{38}$ erg s$^{-1}$.
The best fit, determined by the maximum-likelihood
method, with errors determined with Monte Carlo simulations, gave 
$N_{\circ}=8.7^{+10.4}_{-3.5}$,
$\alpha_{l}=-0.1^{+1.2}_{-4.3}$,
$\alpha_{h}=2.7^{+0.7}_{-0.4}$,
and a break luminosity of
$L_{b}=4.1^{+1.6}_{-1.1} \times 10^{38}$ erg s$^{-1}$.  
Because of the limited range of luminosities observed below the
break luminosity, the low luminosity slope $\alpha_{l}$ is very poorly
determined.

To examine how incompleteness might affect the luminosity function at the
faint end, we simulated the observation using MARX (Model of AXAF Response
to X-rays; Wise et al.\ 2000), version 3.01.
We assumed that the luminosity function of the sources was a single
continuous power-law, which matched the best-fit NGC~1553 luminosity
function above $L_b$.
This power-law was continued to a low luminosity at which the total
LMXB luminosity matched the observed total (resolved and unresolved)
LMXB emission in NGC~1553.
The diffuse emission and background were modeled based on a wavelet
reconstruction of the unresolved emission in NGC~1553, corrected for
unresolved LMXBs.
The sources in the simulated observation were detected using WAVDETECT
using the same criteria as in the real data,
and their count rates were converted into luminosities in the same
way as the real sources.
We found that the luminosity function of the simulated sources was
depressed below the input model luminosity function below a luminosity of
approximately $2 \times 10^{38}$ erg s$^{-1}$.
While this is below the break luminosity determined from the data in
NGC~1553, it shows that incompleteness strongly affects the slope of
the luminosity function below the break luminosity.
Given the large statistical errors in $\alpha_l$ and the substantial
systematic effect of incompleteness, we believe the value of
$\alpha_l$ is unreliable.
Based on the simulation, we do not believe that the break in the observed
luminosity function at $L_b = 4.1 \times 10^{38}$ erg s$^{-1}$ is due
to incompleteness;
however, the systematic uncertainty due to incompleteness increases the
total errors in $L_b$ to about a factor of two.
  
The break luminosity measured for NGC 1553 is about twice the Eddington 
luminosity for
spherical accretion of hydrogen dominated gas onto a 1.4 $M_{\odot}$
neutron star.
Therefore, it may be that the sources with luminosities much higher than
the break are accreting black holes, and those below the break are
predominately neutron stars in binary systems.

A broken power-law was also used to fit the luminosity function of the sources
derived from {\it Chandra} data of NGC~4697
(Sarazin, Irwin, \& Bregman 2000).
The break luminosity found for this galaxy 
($L_{b}=3.2^{+2.0}_{-0.8} \times 10^{38}$ erg s$^{-1}$) is similar to that
of NGC~1553.
The fit to NGC~4697, scaled by the optical 
luminosity (NGC~1553 is 2.04 times brighter than NGC~4697 in the B-band) 
is shown as a dashed line plotted over the luminosity function in
Figure~\ref{fig:lumfunc}.
The cumulative luminosity function of NGC~1553 is higher at the bright
end, and lower at the faint end.
At the bright end, the two luminosity functions agree within the
errors, including the systematic error in the number of background sources.
At the faint end, they disagree significantly.
This is certainly due, at least in part, to incompleteness in the 
luminosity function of NGC~1553 at the faint end.
The luminosity limit for the NGC~4697 observation is
about a factor of four fainter than in NGC~1553.

We determined the $L_{X,src}/L_{B}$ ratio for NGC~1553, where $L_{X,src}$
is the X-ray luminosity in the $0.3 - 10.0$ keV band 
for the resolved and estimated unresolved sources within two effective radii, and 
$L_{B}$ is the B-band luminosity within two effective radii.  We find a 
$L_{X,src}/L_{B}$ value of 
$7.2 \times 10^{29}$ erg s$^{-1}$ $L_{\odot,B}^{-1}$,  
which is slightly lower than the value found for NGC~4697 of
$8.1 \times 10^{29}$ erg s$^{-1}$ $L_{\odot,B}^{-1}$,  
The unresolved source contribution to $L_{X,src}$ was estimated assuming that 24\% of 
the diffuse counts derive from unresolved sources (see \S\ref{sec:RvsD}), based on 
counts in different energy bands and spectral fits.   
For NGC~1553, $L_{X,src}$ is dominated by the unresolved sources, and 
its accuracy is therefore limited by how well we can estimate the contribution to the 
diffuse emission from the unresolved sources.

Another question is whether the number, or total luminosity, of LMXBs
scales with the optical luminosity of the galaxy.
Given that LMXBs represent an old stellar population, it might be that
they scale with the bulge luminosity, rather than the total luminosity,
in S0 galaxies.
An even more intriguing idea is that all or most LMXBs may have formed
in globular clusters (White, Kulkarni, \& Sarazin 2001).
Thus, the number of LMXBs may scale with the number or total luminosity
of globular clusters.
Scaling either with the bulge luminosity or population of globular
clusters would lower the expected number of fainter LMXBs in NGC~1553
relative to NGC~4697, and might explain the observed difference in their  
$L_{X,src}/L_{B}$ ratios.

\section{Resolved versus Diffuse Emission} \label{sec:RvsD}

We have determined the relative contributions of resolved sources and diffuse
emission to the total counts within an elliptical region defined by two 
effective radii.
One effective radius defines the elliptical optical isophote containing one 
half of the optical light; this isophote has a semi-major axis of 78\arcsec, 
an ellipticity of 0.345, and is at a position angle $PA = 149^{\circ}.5$
(Kormendy 1984; Jorgensen, Franx, \& Kjaergaard 1995). 
The two effective radii elliptical region (semi-major axis of 156\arcsec) contains 
all of the emission shown in Figure~\ref{fig:xsmooth}.
In the energy band of 0.3--10.0 keV, the X-ray emission 
is dominated by the diffuse component, which 
makes up 70\% of the counts.  The remaining 30\% of the emission is from the
resolved sources, which are presumably X-ray binaries.  
In the soft S band (0.3--1.0 keV), 84\% of the detected emission is
diffuse, and the remaining 16\% is due
to resolved sources.
In the medium M (1.0--2.0 keV) and hard H (2.0--10.0 keV) bands,
the fractions are
51\% diffuse and 49\% resolved and 
39\% diffuse and 61\% resolved, respectively.
Thus, the diffuse emission is rather soft, while the resolved sources
dominate the hard X-ray band.

The diffuse component 
is much softer than the discrete sources (see \S~\ref{sec:spectra}),
consistent with it being composed at least partly of hot gas rather than
only unresolved X-ray binaries.  Based on the counts from the diffuse component
in the soft (0.3--1.0 keV), medium (1.0--2.0 keV), and hard (2.0--10.0 keV) 
bands we 
estimate that 28\% of the diffuse counts are from unresolved X-ray binaries. 
This assumes that the X-ray binaries make up all of the hard counts.  
Comparing the hard and soft components of the best fit model to the spectrum
of the diffuse emission (\S~\ref{sec:spec_diffuse}), we estimate that hard 
emission makes up
20\% of the counts in the 0.3-10 keV band.  
 From these two methods, then, we estimate that 24$\pm4$\% of the diffuse
counts are a result of unresolved X-ray binaries.

Using the best-fit spectral models for the total emission, resolved
sources, and diffuse emission (\S~\ref{sec:spectra}), we determined 
the unabsorbed flux and
luminosity of each of these components in the band 0.3--10.0 keV.
For the region within two effective radii, the total flux is
$1.43 \times 10^{-12}$ erg cm$^{-2}$ s$^{-1}$.
The total flux from all of the resolved sources, including the bright central
source, within this region is
$5.80 \times 10^{-13}$ erg cm$^{-2}$ s$^{-1}$.
Subtracting the two gives the diffuse flux within two effective radii 
of $8.45 \times 10^{-13}$ erg cm$^{-2}$ s$^{-1}$.  For deriving the fluxes,
note that the spectrum fitted for the sum of the sources includes the central
object and includes the objects within two effective radii only.  This is so
that this flux can be subtracted from the total flux to give directly the
flux of the diffuse emission.  To more carefully examine the spectrum of the 
sources in more detail, in (\S~\ref{sec:spectra}), we sum the sources 
from the entire S3 chip to increase the number of counts in the spectrum, and 
we exclude the central object and object \#38 in an attempt to include only
LMXBs in the spectrum.
Converting the fluxes to
luminosities using a distance to NGC~1553 of 24.2 Mpc gives
$L_{X} = 1.00 \times 10^{41}$ erg s$^{-1}$ for the total emission,
$L_{X} = 4.08 \times 10^{40}$ erg s$^{-1}$ for the sum of the sources, and
$L_{X} = 5.94 \times 10^{40}$ erg s$^{-1}$ for the diffuse emission,
all within two effective radii and for the 0.3 - 10.0 keV energy band.  The
fluxes are all affected by the uncertainty of $Chandra$'s response at low 
energies (below 0.7 keV).  The flux for the diffuse component is most uncertain
because its spectrum is dominated by soft emission.

\section{X-ray Spectra} \label{sec:spectra}

We used the software package kindly provided by Alexey Vikhlinin
(Vikhlinin et al.\ 2000) to extract spectra and to determine the
response matrices.
The response matrices were determined from the FITS Embedded Function (FEF)
response files FP-110\_D1999-09-16fef\_piN0002, appropriate to the
operating temperature of -110 C.
The areas of the detector were weighted by the extracted counts in
determining the response matrices.
For the resolved sources, background spectra were extracted from local regions
around each of the sources.
For the total spectrum or the spectrum of the diffuse emission, we used
a background region as far as possible from NGC~1553 but on the S3 chip
(\S~\ref{sec:data}).
Because the soft X-ray spectral response of the S3 chip was poorly known
at the time when this analysis was done, we restricted our spectral
analysis to the energy range 0.7--10.0 keV.
In order to allow $\chi^2$ statistics to be used, all of the spectra were
grouped to at least 20 counts per spectral bin.
Models were fit to the spectra using XSPEC.
The results are summarized in Table~\ref{tab:spectra},
where the errors are at the 90\% confidence level.

Previous {\it ROSAT} and {\it ASCA} spectra of early-type galaxies have
indicated that they have at least two spectral components,
a very hard component which may be due to X-ray binaries and/or an
AGN
(Matsumoto et al.\ 1997;
Allen et al.\ 2000),
and a softer component.
In X-ray luminous early-type galaxies, the soft component is dominant,
and it is clearly due to diffuse gas at a temperature of $\sim$1 keV
(Forman, Jones, \& Tucker 1985;
Canizares, Fabbiano, \& Trinchieri 1987).
In X-ray faint early-type galaxies, the soft component is much softer,
and its origin is still uncertain
(Fabbiano, Kim, \& Trinchieri 1994;
Pellegrini 1994;
Kim et al.\ 1996;
Irwin \& Sarazin 1998a,b).
The {\it ASCA} spectrum of the hard component has generally been fit by
either a power-law (characterized by a photon spectral index $\Gamma$,
where $\Gamma > 0$ implies a photon spectrum which declines with energy)
(Allen et al.\ 2000)
or by a thermal bremsstrahlung spectrum
(characterized by a hard component temperature $T_h$;
Matsumoto et al.\ 1997).
The soft component in X-ray bright galaxies is usually fit by
the Raymond-Smith (Raymond \& Smith 1977) or MEKAL model for the emission
from a low density, optically thin
plasma
(Fabbiano et al.\ 1994).
This model is characterized by the temperature of the gas ($T_s$)
and by the abundances of the heavy elements.
Given the limited statistics we have in our spectra, we will assume
that the heavy element abundances have the solar ratios, and only allow
the overall abundance of the heavy elements to vary.
In X-ray faint galaxies, it is unclear what the appropriate soft emission
model should be.
If the soft emission is due to diffuse gas, then the MEKAL model would
again be appropriate.
If it is due to an optically thick stellar component (including the
same LMXBs which produce the hard component), then it might be
better represented as a blackbody, characterized by a temperature ($T_s$
again).
Thus, we have used spectral models which include both a hard (power-law
or bremsstrahlung) and soft (MEKAL or blackbody) component.

\subsection{Total X-ray Spectrum} \label{sec:spec_total}

\begin{figure}[t!]
\vskip 5.0truein
\includegraphics{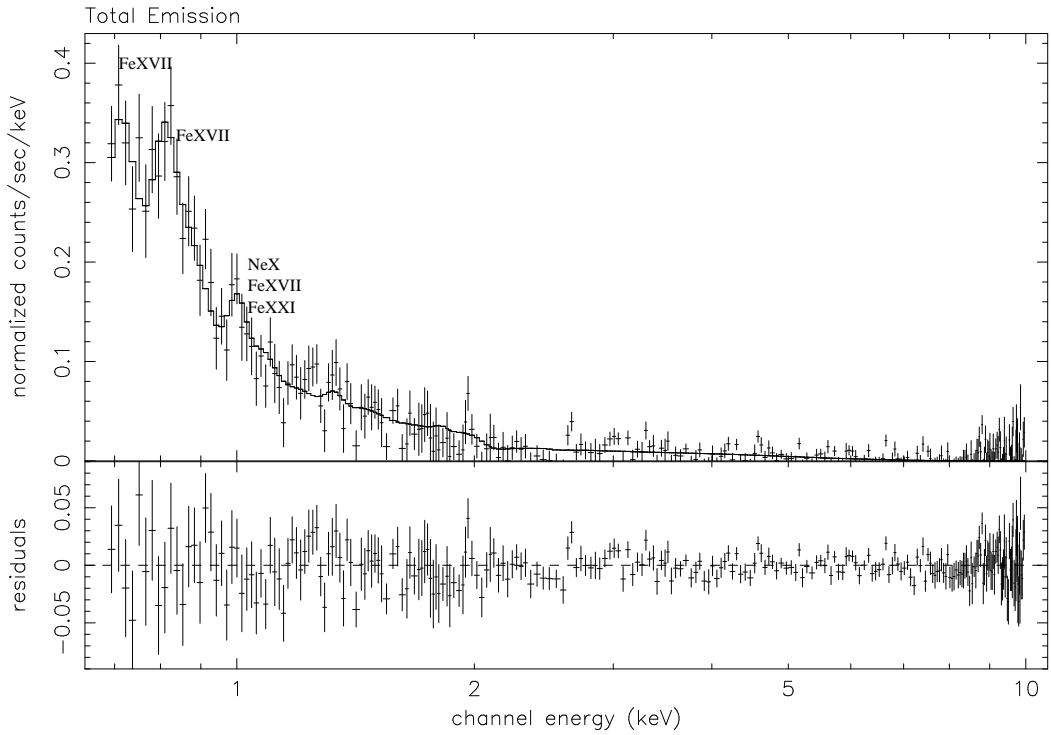}
\caption{X-ray spectrum of the total emission (diffuse plus resolved sources)
from within two effective radii.
The points with error bars in the upper panel are the data, while the
solid histogram is the best fit model including Galactic absorption,
a MEKAL soft emission component, and a power-law hard component.
The lower panel shows the residuals to the model.
\label{fig:spec_total}}
\end{figure}

We first determined the total spectrum (resolved sources and diffuse
emission) from within two effective radii.
The spectrum is shown in Figure~\ref{fig:spec_total}.
We tried fitting the spectrum with either a pure hard component
(power-law or bremsstrahlung) or a pure soft component (MEKAL or
blackbody), but they did not provide an adequate fit.
We allowed the absorbing column $N_H$ to vary, and found that the fit was
not significantly improved ($\Delta\chi^{2} < 1$) over using the Galactic 
column towards NGC~1553 of $1.41 \times 10^{20}$ cm$^{-2}$
(Dickey \& Lockman 1990).  We therefore fixed the absorption to the Galactic 
value.

When the hard component is modeled by thermal bremsstrahlung, 
the temperature diverges and is unconstrained
(XSPEC imposes a limit of $kT_h \le 200$ keV).
At such high temperatures,
the shape of thermal bremsstrahlung at lower energies resembles a
power-law with a photon index of $\Gamma \approx 1.4$.
This suggests that a power-law might provide a better fit for the
hard component.
Indeed this is found to be the case, although the improvement is
not very substantial.  We list two fits in Table~\ref{tab:spectra}, one with
the photon index a free parameter, and the other with it fixed to the value
found from fitting the sources (\S\ref{sec:spec_sources}) of 
$\Gamma =  1.20$ since we expect that it is discrete sources that make up 
the hard component.  The fit with the free photon index is better in the
0.7--10.0 keV range, however, when
we expand the energy range to 0.3--10.0 keV and do not refit the model, 
the fit with the fixed photon index is much better.
Therefore, we use the model with the fixed photon index when determining 
the flux and luminosity in the 0.3--10.0 keV range 
(the same is true for the diffuse, spiral, and circle components 
described below).

We tried to fit the soft component with a MEKAL, blackbody (bbody)
or disk blackbody (diskbb) model.
The MEKAL model provided the best fit.
The temperature of the cooler component was $\sim$0.47 keV,
and the abundance was $\sim$0.16.  
This temperature is consistent with that determined from the
temperature vs.\ velocity dispersion ($kT-\sigma$)
correlation derived from $ROSAT$ observations of elliptical galaxies 
(Davis \& White 1996), using a velocity dispersion for NGC~1553 of 
$\sigma=184$ km s$^{-1}$ (Longo et al.\ 1994).  The abundance is consistent
with the temperature-abundance relation reported by Davis \& White (1996),
who found that abundance increases as temperature increases.
The spectrum in the 0.7--10.0 keV range is displayed in Figure~\ref{fig:spec_total} with the model including absorption fixed at the Galactic value, a MEKAL
component, and a power-law with the photon index set to 1.20.
The observed spectrum clearly shows the (mainly)
\ion{Fe}{17} ($\sim$0.72 keV emitted),
\ion{Fe}{17} (0.826 keV emitted),
and the
\ion{Ne}{10}, \ion{Fe}{17}, \ion{Fe}{21} ($\sim$1.02 keV)
line complexes,
indicating that a significant portion of the soft emission is thermal
emission from gas.
For this fit
the luminosities of the hard and soft components are
$L_{X,h} = 7.79 \times 10^{40}$ erg s$^{-1}$ and
$L_{X,s} = 2.23 \times 10^{40}$ erg s$^{-1}$, respectively, giving a ratio
$L_{X,h}/L_{X,s} = 3.45$.  
The ratio of the hard-to-soft model count rates is 0.99.

We simultaneously fit the $Chandra$ spectrum of the total emission with a 
spectrum extracted from an identical region in the $ROSAT$ PSPC using a
model combining  MEKAL and power-law components.  The 
resulting fit agreed with the fits to the total emission described in
Table 2 within the errors.

Trinchieri et al.\ (2000) fit a two-component model to a spectrum derived 
from $BeppoSAX$ MECS and LECS data of NGC~1553.  Using a Raymond-Smith model, 
they found a lower temperature ($kT = 0.26$ keV) for the soft component,
although the two measurements overlap within the errors.  The LECS is sensitive
to lower energies (down to 0.1 keV) than our cutoff of 0.7 keV, and sampling
this low energy range has a significant effect on determining the temperature
of the soft emission.  In addition, Trinchieri et al.\ extracted their spectrum 
from a larger spatial region than we did.  If there is a negative temperature
gradient, we would expect the average temperature derived from the spectrum
from this larger region to be lower. 
Trinchieri et al.\ used a thermal bremsstrahlung model to fit the hard emission
component.  They found a temperature of 4.8 keV for the hard component which
therefore has a steeper decline with photon energy than we found.

\subsection{Diffuse X-ray Spectrum} \label{sec:spec_diffuse}

We extracted the spectrum of the diffuse emission (excluding the
resolved sources) from within two effective radii.  Again, we tried fitting
several models.  The best fit model required both a soft MEKAL component and 
a hard power-law component.  
This fit is better than a fit of just a soft component
at the 99.93\% confidence level, and better than the fit of a hard component
model at the 99.99\% level. The absorption was set to the Galactic value 
since freeing it did not improve the fit.  Two fits are listed in 
Table~\ref{tab:spectra}, one with the photon index allowed to vary and one
with it fixed to the source value of $\Gamma = 1.20$.  As with the total
emission, the fit is better in the 0.7--10.0 keV range when the photon index
is free.  Expanding the energy range to 0.3--10.0 keV without refitting the
model shows that the model with the fixed photon index is much better at low
energies.

\begin{figure}[t!]
\vskip 5.0truein
\includegraphics{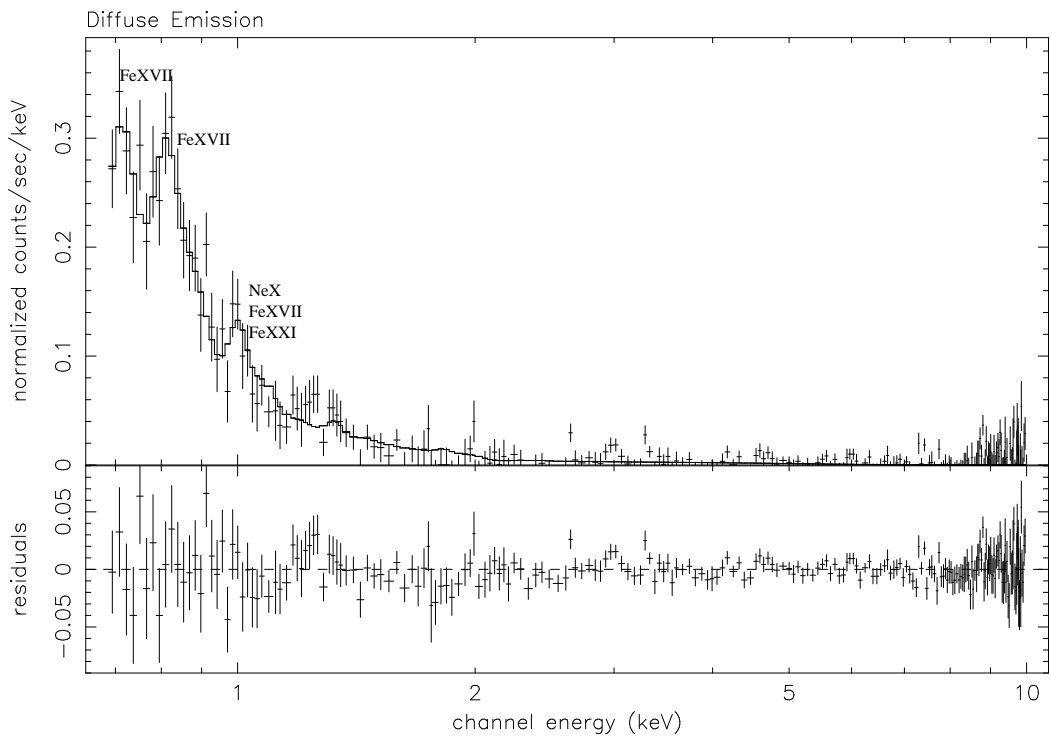}
\caption{Spectrum of the diffuse emission within two effective radii fit with
a model combining Galactic absorption, a MEKAL soft component, and a
hard power-law hard component.
\label{fig:spec_diffuse}}
\end{figure}

The spectrum of the diffuse component is shown in 
Figure~\ref{fig:spec_diffuse}, along with a solid line representing a model
including Galactic absorption, MEKAL emission, and a power-law with the photon
index fixed to the source value.  
The ratio of the hard-to-soft model count rates is 0.26.
The ratio of the luminosities of the hard to soft components in the
diffuse emission is $L_{X,h}/L_{X,s} = 0.88$, 
which is smaller than the value for the
total emission ($L_{X,h}/L_{X,s} = 3.45$), or the resolved sources 
($L_{X,h}/L_{X,s} = 6.85$, see below).
Thus, the diffuse emission is mainly soft X-rays.
The soft spectrum and the good fit to a MEKAL model suggest that
much of the soft emission is due to diffuse gas.
As was true to the total spectrum, the diffuse spectrum shows the
(mainly)
\ion{Fe}{17} ($\sim$0.72 keV emitted),
\ion{Fe}{17} (0.826 keV emitted),
and the
\ion{Ne}{10}, \ion{Fe}{17}, \ion{Fe}{21} ($\sim$1.02 keV)
line complexes,
which supports the predominance of thermal emission due to hot gas.
The hard component in the diffuse emission may be due to unresolved
LMXBs below the limit of detection as discrete sources.

\subsection{X-ray Spectrum of Resolved Sources} \label{sec:spec_sources}

We determined the spectrum of the sum of all of the resolved sources on the
S3 chip,
excluding the central source and source \#38.  Src.~38
is likely an unrelated background source and we therefore exclude
it when extracting the spectrum of the sum of the sources.  Several pieces of
evidence point towards the central source being an AGN rather than an LMXB like
most of the other discrete sources, so we analyze its spectrum separately in 
\S~\ref{sec:spec_agn}.

\begin{figure}[t!]
\vskip 5.0truein
\includegraphics{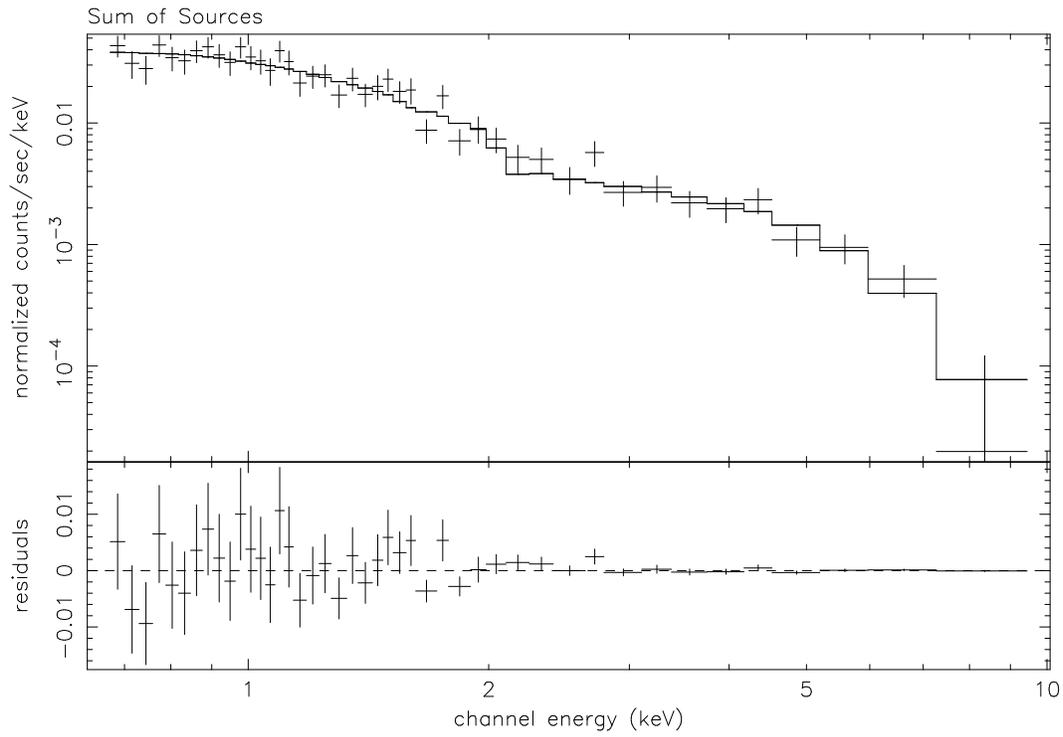}
\caption{Spectrum extracted from a sum of the sources, excluding the central 
source and the bright Src.~38 fit with
a model combining Galactic absorption, soft blackbody emission, and a
hard power-law.
\label{fig:spec_sources}}
\end{figure}

We achieved the best fit to the spectrum by using a model combining 
a soft, blackbody component, and a hard, power-law component.
The f-test showed
that the fit including both a soft and hard component was better than a fit 
with just a hard, power-law component at a significance level of 96\%.
When a MEKAL model was used for the soft component, the abundance and
temperature were very poorly determined, and there is no real evidence
for line emission in the spectrum.
Thus, we also fit the soft component with a blackbody spectrum
which gave an acceptable fit.  
The fit was not improved by allowing the absorption to vary so we set it
to the Galactic value.
The spectrum is shown in Figure~\ref{fig:spec_sources},
along with the fit of a model combining absorption set to the Galactic value,
blackbody emission, and a power-law component.  The best fitting model gave
a blackbody temperature of $kT_{s} = 0.24$ keV and a power-law photon index of
$\Gamma = 1.20$.

The resolved sources have a harder spectrum than that of the diffuse
emission.  
The ratio of the hard-to-soft model count rates is 2.08.
The luminosities of the hard and soft components, excluding 
the central source and Src.~38, are
$L_{X,h} = 2.35 \times 10^{40}$ erg s$^{-1}$ and
$L_{X,s} = 3.43 \times 10^{39}$ erg s$^{-1}$, respectively, giving a ratio
of $L_{X,h}/L_{X,s} = 6.85$. 

\subsection{X-ray Spectrum of Central Source} \label{sec:spec_agn}

\begin{figure}[t!]
\vskip 5.0truein
\includegraphics{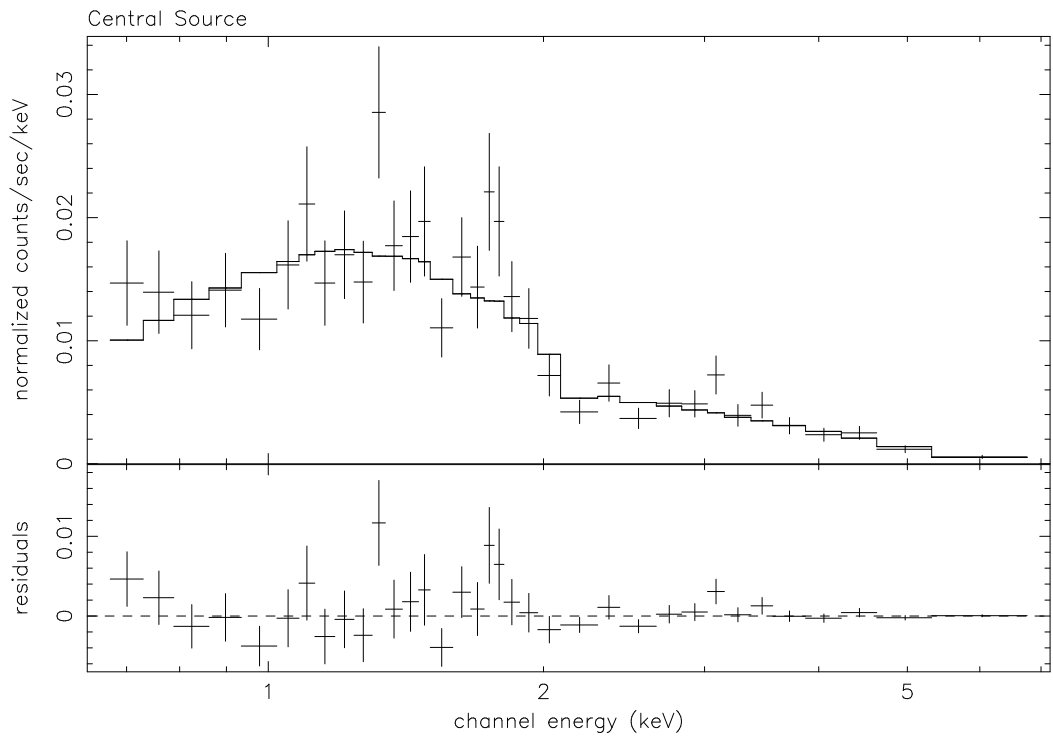}
\caption{Spectrum of the bright, central resolved source fit with
a model combining a high absorbing column and disk blackbody emission.
\label{fig:spec_agn}}
\end{figure}

We determined the X-ray spectrum of the central X-ray source, which
is the brightest source in the field.
The spectrum and best fit model are shown in Figure~\ref{fig:spec_agn}.
This spectrum is best fitted with a disk blackbody model with a temperature of
$kT = 1.70$ keV and a high 
absorbing column ($18.25^{+7.42}_{-7.59} \times 10^{20}$ cm$^{-2}$), much higher than Galactic.
This hard, absorbed spectrum is consistent with the hardness ratios
of this source (\S~\ref{sec:hardness}).
This spectrum indicates that the central source is different than most of
the other resolved sources, which are presumably LMXBs.
The hard spectrum and high absorption suggest that this source is
an obscured AGN.  
The disk blackbody model provides a better fit than a power-law model at
the 66\% level as shown by the f-test.  The power-law model
has a best-fitting photon index of $\Gamma=1.59$, and a high absorbing
column of $36.04^{+12.96}_{-9.22} \times 10^{20}$ cm$^{-2}$).  This photon 
index is consistent with that seen for Weak Line Radio Galaxies and LINERs (e.g.,
Sambruna et al. 1999).

\subsection{X-ray Spectrum of Spiral Feature} \label{sec:spec_spiral}

\begin{figure}[t!]
\plotone{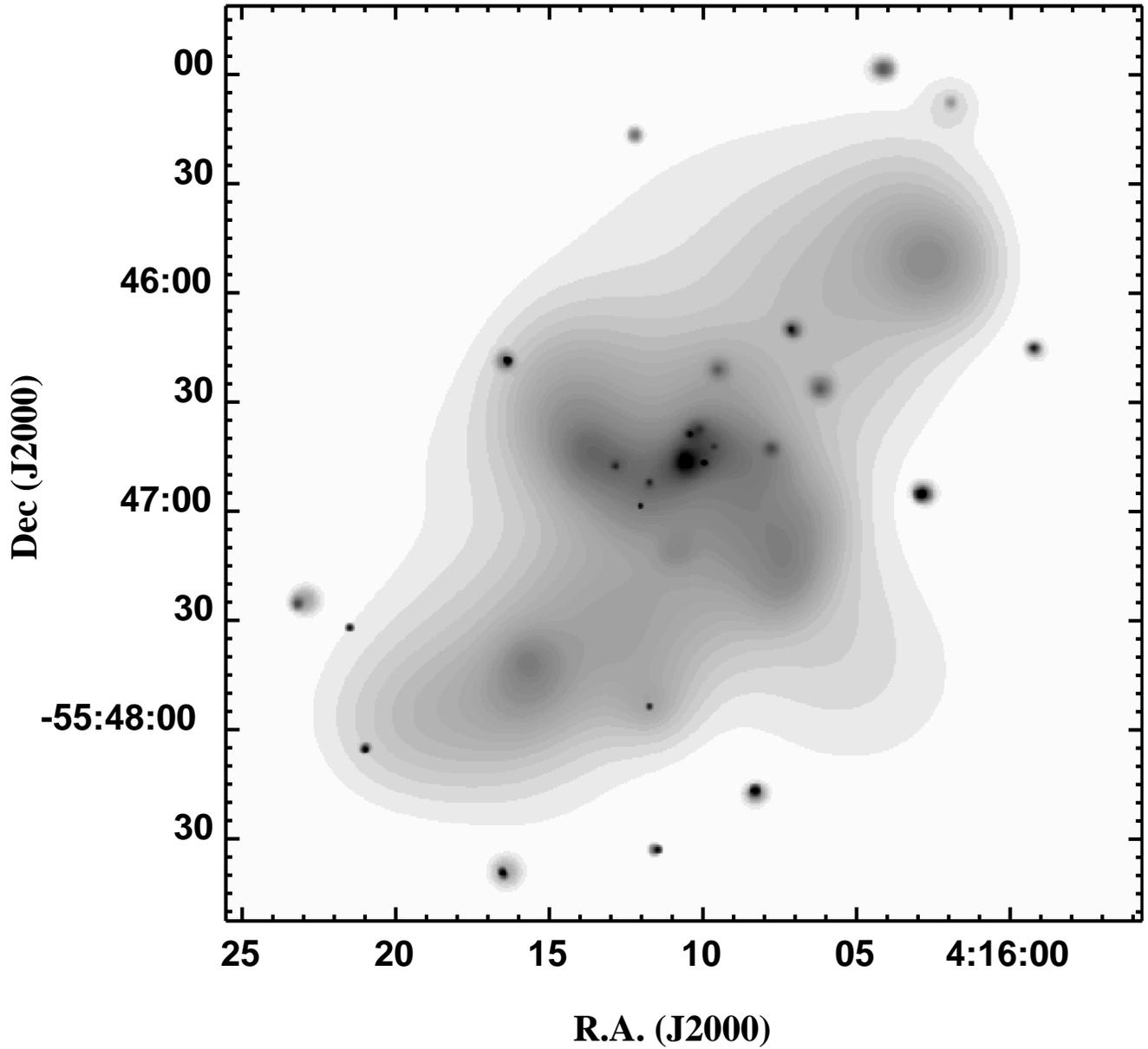}
\caption{Adaptively smoothed image of the $4^{\prime}\times4^{\prime}$ region
surrounding the center of NGC~1553.  The image has been corrected for exposure
and background.
The grayscale is 
logarithmic and ranges from $3.8\times10^{-7}$ to $3.0\times10^{-3}$ ct 
pix$^{-1}$ s$^{-1}$.
\label{fig:xsmooth}}
\end{figure}

In order to study the origin of the spiral feature
seen in the diffuse emission (Figs.~\ref{fig:xwhole}, \ref{fig:xsmooth}),
we determined its spectrum.
A spatial region was constructed which covered the feature as seen
in Figure~\ref{fig:xsmooth}, but excluding the central source and other
resolved sources.
In order to compare this feature with the surrounding diffuse emission,
we also extracted the spectrum for a circular region centered on the
galaxy that just enclosed the spiral region, and excluded the spiral region
and discrete sources found within the circle.

Best fitting models to the
spiral and circular regions are listed in Table~\ref{tab:spectra}.
As above, letting the absorption vary did not improve the fits, so we set 
it to its Galactic value.  When the photon indices are allowed to vary, the
spiral component has a best fitting power-law that is inverted with 
$\Gamma = -1.55$.
Examination of the spectrum indicates that there is excess
hard emission coming from the spiral region.  This emission probably also 
affects the fits to the total and diffuse components which both include the
spiral region and also have best
fitting power-laws with low or negative photon indices.  The circular region
which excludes the spiral emission has a photon index consistent with 
that of the sources.  The high abundance indicated by the fit of the spiral
component with the photon index fixed to the source value is the result of 
trying to fit the lines at high energies where the model continuum is 
lower than in
the fit with the inverted power-law.  This high abundance, therefore, is
most likely not real.  The abundance measured with the fit using the free
photon index is consistent with the abundance measured for the circular, total,
and diffuse regions.
The excess hard emission in the spiral region may be
non-thermal, inverse Compton radiation resulting from a shock.  A 
high-resolution radio map would be useful in determining more definitively
the origin of this emission.

\section{Discussion} \label{sec:disc}

\subsection{Nature of the Central Source} \label{sec:agn}

In principle, the central X-ray source in NGC~1553 might be due to
one or several LMXBs and/or to a central AGN.
However, previous observations in other wavebands indicate that it is an AGN,
and our X-ray observations support this picture.
NGC~1553 is a weak radio source (10 mJy at 843 MHz; Harnett 1987),
with a LINER-like optical emission spectrum (Phillips et al.\ 1986).
The X-ray evidence that suggests the central source is an AGN follows.
First, the position of this source agrees with the position of the
optical nucleus of the galaxy to within the errors.
Second, its luminosity is $L_X = 1.75 \times 10^{40}$ erg s$^{-1}$, which
would make it an extremely luminous X-ray binary.
This luminosity corresponds to the Eddington limit for spherical accretion
by a $\approx$130 $M_\odot$ compact object.
While such luminous, high mass, presumably black hole binaries may have
been found in star formation regions in spiral galaxies
(e.g., Kaaret et al.\ 2000),
it seems unlikely that they would occur in LMXBs.

Third, the central source has a harder X-ray spectrum than is typical
of the other resolved sources in this galaxy,
as shown both by its hardness ratios (\S~\ref{sec:hardness}) and its
detailed X-ray spectrum (\S~\ref{sec:spec_agn}).
The X-ray spectrum has very strong soft X-ray absorption, which is
completely inconsistent with Galactic column in this direction
(Table~\ref{tab:spectra}).
The X-ray spectrum is consistent with that expected for an obscured
AGN, surrounded by a significant column of nuclear gas.

\subsection{Relation of Diffuse Emission Features to Cooler Gas}
\label{sec:halpha}

In addition to the hot, X-ray emitting gas, NGC~1553 contains cooler
gas seen in optical emission lines and as dust extinction features
(Phillips et al.\ 1986;
Bettoni \& Buson 1987;
Bally \& Thronson 1989;
Roberts et al.\ 1991;
Trinchieri et al.\ 1997;
Quillen et al.\ 2000). 
Based on a lower spatial resolution {\it ROSAT} HRI X-ray image,
Trinchieri et al.\ (1997) argued for an association of a region
SE of the nucleus of NGC~1553 seen in H$\alpha$ emission
with a region of enhanced X-ray emission.
The X-ray region noted by Trinchieri et al.\ is apparently the extension
to the SE in Figure~\ref{fig:xsmooth}.
We agree with this association of the X-ray emission with H$\alpha$
emission.
On the other hand, there is no corresponding H$\alpha$ feature associated
with the extension to the NW, and, on smaller scales near the center of
the galaxy, H$\alpha$ emission is not aligned with the ``spiral feature.''
A north-south dust lane, approximately 3$^{\prime\prime}$ across, is seen
with HST at the center of the 
galaxy (Quillen et al.\ 2000).
It runs approximately perpendicular to the spiral region, and has 
no correlation with any of the X-ray emission features.  The position angle
of the dust lane seems to approximately agree with that of small-scale 
H$\alpha$ emission at the center of the galaxy (Trinchieri et al.\ 1997).
Overall, there isn't much evidence for a strong correlation of the
X-ray morphology with cooler gas and dust.

\subsection{Origin of the Spiral Feature} \label{sec:spiral}

We now consider the origin of the ``spiral feature'' seen in X-rays
near the center of NGC~1553 (Figs.~\ref{fig:xwhole} \& \ref{fig:xsmooth}).
The hardness ratios, or ``colors''
for these regions are indistinguishable from those of the diffuse
emission as a whole, and the emission is predominately soft
(approximately 70\% of the counts fall in the 0.3--1.0 keV energy band).
Spectral fits (\S~\ref{sec:spec_spiral}) show that the gas forming the
spiral feature has approximately the same temperature,
abundance, and absorbing column as the surrounding gas. 
We also determined the average X-ray surface brightness enhancement of the
spiral feature compared to the surrounding diffuse emission, using a number of
regions to collect counts from the feature and from a circular region
which just enclosed it.
In the 0.3--10 keV band, the spiral feature has an average surface brightness
enhancement of 1.82 compared with the emission immediately surrounding it.
We converted this surface brightness contrast into a density contrast,
assuming the best-fit spectra for the spiral and surrounding gas, and
for two simple geometries.
First, we assume that the extent of the spiral feature along the line of
sight is comparable to its width (e.g., it is a narrow filament) on
the plane of the sky (an angle of approximately 35$^{\circ}$).
In this case, the spiral gas is denser than the surrounding material by
an average factor of $\rho_{sp} / \rho = 1.87$.
As an alternative, we assume that the extent of the spiral feature along
the line of sight is comparable to the distance from the nucleus of
the galaxy
(e.g., it is a sheet of enhanced emission).
This sheet geometry may require that we are observing the system at
a preferred angle (along the sheet), or that it be part of a limb-brightened
structure.
For this geometry, the density contrast of spiral gas is an average
factor of about $\rho_{sp} / \rho =1.35$.

First, we consider the possibility that this feature is due to absorption;
that is, the spiral is not an emission feature, but the result of excess
absorption in the surrounding region.
In fact, the spectral fits indicate that there is not enough excess
absorption in the surrounding region to provide a significant surface
brightness enhancement for the spiral.
Also, if the required excess absorption were due to cooler gas, one might
expect the spiral feature to be surrounded by optical emission lines or
dust absorption;
neither is the case (\S~\ref{sec:halpha}).
It also seems improbable that absorption in the surrounding gas would
produce such a well defined spiral feature.

Second, X-ray enhancements might be due to the interaction between hot
X-ray emitting gas and cooling material, leading to the evaporation of the
cooler material
(Kim 1989;
Sparks, Macchetto, \& Golombek 1989;
Macchetto et al.\ 1996;
Sparks 1997).
This model for the spiral feature would require that there be a very close
association with optical emission line gas and/or dust.
As noted above (\S~\ref{sec:halpha}), no such association is seen
in NGC~1553.

Third, the spiral features might be tidal in origin.
In support of this idea, we note that NGC~1553 is interacting
with the companion elliptical galaxy NGC~1549.
There are shell or tidal features seen in NGC~1553
(Malin \& Carter 1983; Bridges \& Hanes 1990), but all are at large radii.
There are no distortions apparent in the HST optical image of NGC~1553
which are associated with the spiral feature, or which are located
at similar radii.
If the spiral feature was due to gravitational tides, one would expect
the stars to show a similar effect.

Fourth, the spiral might be due to gas which is radiatively cooling, and
then being sheared as a result of slow rotation.
In X-ray bright early-type galaxies, the cooling time is short in
the central regions, and the gas is believed to form a cooling flow
(e.g., Sarazin 1990).
We estimate the radiative cooling time of gas in the spiral region.
Using the peak 
surface brightness of the spiral region of $7.5 \times 10^{-6}$ ct pixel$^{-1}$
s$^{-1}$, along with the best fitting spectral model, we determine the 
approximate density of this region for both the filamentary and sheet-like 
geometries.
For the filamentary (sheet) geometry, we estimate the density
corresponding to the peak in surface brightness
as $n_{e} = 0.026$ (0.019) cm$^{-3}$.
The isobaric cooling time is then 
$t_{cool} \approx 1.5~(2.0) \times 10^{8}$ yr for the filamentary (sheet)
geometry.
This is much shorter than the probable age of the system, so it is likely
that the gas in the spiral is cooling, unless some other source of energy
is replacing the radiative losses.

In this case, we would expect the gas in the spiral feature to be cooler
than the surrounding gas.
The spiral component has a best fitting MEKAL temperature of
0.51$^{+0.07}_{-0.08}$ keV and that of the
circular region is 0.48$^{+0.07}_{-0.09}$ keV.
The ratio of the temperature in the spiral region to the temperature of the 
surrounding region is then
$T_{sp}/T = 1.05^{+0.23}_{-0.23}$ at a 90\% confidence level.
Obviously, this doesn't rule out the spiral gas being cooler than
the ambient gas.
On the other hand, the radiative cooling time is much longer than the
dynamical time for the gas, $t_{dyn} \approx 10^{7}$.
Thus, one would expect the cooling gas to be nearly in
pressure equilibrium with the surrounding.
If we take the lower limit of the temperature contrast
of $T_{sp}/T > 0.81$, and use density contrasts derived above, we find
a minimum pressure ratio of $P_{s}/P > 1.10$ for the long (sheet)
geometry and 
$P_{s}/P > 1.52$ for the narrow (filament) geometry.
Other sources of pressure support, such as magnetic fields, would only
increase this limit further.
Thus, it seems unlikely that the gas in the spiral feature in
in pressure equilibrium.

Another concern is the large contrast between the cooling time and
the dynamical time in the gas.
It seems unlikely that the spiral structure could be maintained while
the gas cooled.
This would probably require a very small but nonzero rotation and
some other mechanism, such as magnetic fields, to stabilize the
cool gas regions against infall or being sheared away.
It is also difficult to understand why the feature would be two armed
and relatively symmetric if it were due to radiative cooling of
random perturbations in the ambient gas.
The sum of the evidence suggests that it is unlikely that the spiral
feature is due to cooling.

Next, we consider the possibility that the spiral region is due to
adiabatic compression;
If we ignore magnetic fields or other pressure sources,
adiabatic compression requires that
$T_{sp}/T = ( \rho_{sp} / \rho )^{2/3}$.
For the density contrast associated with a filamentary (sheet-like) geometry,
this implies a temperature contrast of $T_{sp} / T = 1.52$ (1.22).
The observed temperature contrast is consistent with adiabatic compression
for a sheet-like geometry, but not for the filamentary geometry.

Finally, we consider the possibility that the spiral features are
shocks.
Again, we assume an ideal gas, with no magnetic fields or
other sources of pressure.
Then, the temperature increase $T_{sp}/T$ and shock compression
$r \equiv \rho_{s}/\rho$ are related by
(Markevitch, Sarazin, \& Vikhlinin 1999)
\begin{eqnarray}
\frac{1}{r} =
\left[\frac{1}{4}\left(\frac{\gamma+1}{\gamma-1}\right)^2
\left(\frac{T_{sp}}{T}-1\right)^2 +\frac{T_{sp}}{T}\right]^{1/2} \cr
-\frac{1}{2}\frac{\gamma+1}{\gamma-1}\left(\frac{T_{sp}}{T}-1\right),
\label{eq:shock1}
\end{eqnarray}
with $\gamma = 5/3$.
 From the spectrally-derived upper limit on the temperature contrast of
$T_{sp}/T < 1.282$, 
we find an upper limit to the compression
$r = \rho_{sp}/\rho < 1.427$.
Therefore, a simple shock model using a sheet-like geometry can
explain the increase in surface brightness, although a filamentary
geometry cannot.
The velocity discontinuity $\Delta u_{sh}$ at the shock is given by
(Markevitch et al.\ 1999)
\begin{equation}
\Delta u_{sh} = \left[\frac{k T}{\mu m_p}\left(r-1\right)\left(
\frac{T_{sp}}{T}-\frac{1}{r}\right) ,
\right]^{1/2} \, , \label{eq:shock2}
\end{equation}
where $\mu = 0.61$  is the mean mass per particle in terms of the mass
of the proton $m_p$.
For the upper limit temperature increase, the velocity change at the
shock is $\Delta u_{sh} < 138$ km s$^{-1}$, and the shock velocity relative
to the ambient gas is
$u_{sh} = \Delta v_{sh} r / ( r - 1 ) < 462$ km s$^{-1}$.

Thus, it seems most likely that the spiral feature results from subsonic
or moderately transonic compression of the ambient gas in the
region.
The gas is required to be in a sheet rather than a filament.
The required velocities are comparable to orbital velocities in the
galaxy, so it is possible that the spiral feature is a shell,
similar to the stellar dynamical shell seen in the optical in the
outer part of NGC~1553.

On the other hand, the spiral feature might be a part of a limb-brightened
shell of gas compressed by expanding radio lobes from the radio nucleus
(Harnett 1987).
Since the lifetimes of radio sources (typically, $\la 10^7$ yr)
are shorter than the cooling time of the gas,
this model can explain why the gas has been compressed without significant
radiative cooling.
Similar shells have been seen in several cooling flow cluster central galaxies
(McNamara et al.\ 2000;
Fabian et al.\ 2000).
This would explain the nearly symmetric geometry
(i.e., why it is present on both sides of the nucleus).
The required radio jets and lobes would extended in a direction
perpendicular to the dust and H$\alpha$ disk near the center of the
galaxy
(Trinchieri et al.\ 1997;
Quillen et al.\ 2000).
The geometry has been seen in other radio galaxies
(e.g., McNamara et al.\ 2000), and is consistent with the dust lane being
the outer portion of an accretion torus around the AGN, with the radio
jets emerging along the rotation axis of the torus.
This model predicts that the spirals form as rims around the radio source,
which must have a double jet/lobe structure.
The S-shaped X-ray structure may indicate that the radio jets are
not propagating along a symmetry axis of the galaxy, as it also
required if the jets are perpendicular to the dust lane.
In this case, each of the jets would encounter denser gas on one side,
and an S-shaped structure would be produced.

Unfortunately, we have been unable to find a radio image of this
southern galaxy which resolves the galaxy.
The only information available is the approximate flux density of
10 mJy at 843 MHz
(Harnett 1987).
Upper limits at higher frequencies do not usefully limit the radio
spectral index.  There are several other published radio flux densities for 
this galaxy, however, as revealed by the Harnett (1987) map, they include an
unrelated bright neighboring source in the flux density measurements.  
The radio source associated with the nucleus of NGC~1553 is
rather faint, and may be too weak to affect the
hot interstellar gas in NGC~1553 enough to produce the spiral features.
To test very crudely this idea, we assume that the radio source has
displaced two spherical regions of thermal gas, each with a diameter of
30\arcsec.
We assume that the minimum pressure in the radio source is the estimated
thermal pressure in the spiral feature assuming the ``sheet'' geometry,
$P_{sp} \approx 3 \times 10^{-11}$ dynes cm$^{-2}$.
Then, the minimum total energy content of the two radio lobes is
$E_{radio} \ga 6 \times 10^{55}$ ergs.
The radio power at a frequency of $\nu = 847$ MHz is
$P_\nu \approx 7 \times 10^{20}$ W Hz$^{-1}$.
If we assume that the radio spectrum is a power-law
$P_\nu \propto \nu^{\alpha}$ extending from 0.01 to 100 GHz, then the total
luminosity is $L_{radio} \approx ( 5 - 11 ) \times 10^{37}$ ergs s$^{-1}$ for
$-1.5 < \alpha < -0.7$.
We assume that the radio source is powered by a jet, and that the
efficiency of conversion of jet power $L_{jet}$ into radio luminosity is
$\epsilon_{jet}$, so that
$L_{jet} \approx ( 5 - 11 ) \times 10^{39} ( \epsilon_{jet} / 0.01 )^{-1}$
ergs s$^{-1}$.
In order to inflate the required cavities, this jet would have to
operate for a time scale of
$t_{radio} \approx E_{radio} / L_{jet} \ga 2 \times 10^8 ( \epsilon_{jet} /
0.01 )$ yr.
This is somewhat longer than the ages usually assumed for radio sources,
so either the jet efficiency would need to be unusually low, or the
radio source is presently weak but was stronger in the recent past.

As a alternative test of the radio interaction hypothesis for the spiral
feature, we assume that the observed radio flux comes from the two
radio cavities needed to produce the structure, and determine the
minimum energy content of these lobes.
In addition to the previous assumptions, we assume that the magnetic
field is perpendicular to the line of sight and that the ratio
of energy in ions to electrons in the radio lobes is unity.
Since the radio spectral index is unknown, we allow it to vary and
require that the minimum radio source energy exceed that required to
inflate the cavities.
This requires that $\alpha \la -2.7$, which is an extremely steep spectrum.
Instead, if we assume the ratio of ions to electrons is 100, the
limit is $\alpha \la -1.7$.
These results indicate that the radio source can explain the spiral feature
only if it is far from the minimum energy or equipartition limit.
Both of these arguments suggest that the spiral feature can be explained
by an interaction with the radio source only if the radio source is
a fading remnant of previously more intense radio activity. 

\section{Conclusions}

Using a $Chandra$ observation of the X-ray faint S0 galaxy NGC~1553, we have
spatially and spectrally resolved the sources of its X-ray emission.  A
significant fraction, approximately 70\%, of the total X-ray counts in the
0.3--10.0 keV band are detected
as diffuse emission and 49 discrete sources are resolved.
The diffuse emission dominates at soft energies, making up 84\% of the counts
from 0.3--1.0 keV, while the resolved sources make up most of the hard
emission with 61\% of the counts in the 2.0--10.0 keV band.  We estimate that
approximately 24\% of the counts from the diffuse component derive from
unresolved LMXBs.
The diffuse component is asymmetric, with extensions to 
the NW and SE, as well as a spiral-like structure running SW to NE.  These
asymmetries are consistent with the diffuse emission arising at least
partly from hot gas rather than only unresolved LMXBs.

Three of the resolved sources are coincident with globular clusters found with
HST, and an additional six have apparent optical associations on the DSS. 
The luminosities of the individual sources range from $1.9 \times 10^{38}$ to
$1.8 \times 10^{40}$ erg s$^{-1}$, with the central source being the brightest.
The luminosity function of the resolved sources is well fit by a 
broken power-law model with a break radius of approximately twice the 
Eddington luminosity for spherical accretion of hydrogen dominated gas onto a
$1.4 M_{\odot}$ neutron star.  Those sources with luminosities above the 
break are likely accreting black holes, and those with $L_X$ below the break
are probably mainly neutron stars in binary systems.
The luminosity function and break luminosity are similar to that seen
with {\it Chandra} in the elliptical NGC~4697
(Sarazin et al.\ 2000).
Thus, the break luminosity make be a general feature of the LMXB
populations of early-type galaxies.
The break luminosity could be used as a distance indicator,
although it is unlikely to be competitive with other techniques,
given the large errors associated with the relatively small number
of bright LMXBs.
In any case, these LMXBs provide a new window on the stellar evolution
of early-type galaxies, which is particularly useful for studying the
massive stars which were present in these galaxies long ago but which
have long since vanished from the optical band.

The composite X-ray spectrum of the resolved sources is best fit with a two
component model combining soft blackbody emission with $kT = 0.24$ keV,
and a hard, power-law component with a photon index $\Gamma = 1.20$.
The luminosity of the sources is dominated by the hard emission.
The weakness of the soft emission from the LMXBs suggests that it
is unlikely that they dominate the soft emission seen in the
X-ray faint galaxies
(Fabbiano, Kim, \& Trinchieri 1994)
as has been suggested by two of us
(Irwin \& Sarazin 1998a,b).

The strongest source in the field is located at the center of NGC~1553
to within 0\farcs5.
The luminosity ($1.75 \times 10^{40}$ erg s$^{-1}$)
of this source is rather high for an LMXB.
The spectrum of this source is very different than that of the LMXBs in
the galaxy.
It has a hard spectrum with a very high absorbing column.
This adds to the evidence that the central source is due to the central
AGN in this galaxy, which is a radio source
(Harnett 1987)
with a LINER-like optical emission spectrum
(Phillips et al.\ 1986).

For the X-ray spectrum of the diffuse emission and of the total (sources
plus diffuse) emission, the best fit was achieved with a model combining
a soft MEKAL component, with a temperature $kT \approx 0.5$ keV and abundance 
approximately 10\% solar, along with a hard, power-law component.
In this case, the soft component is dominant.
The total and diffuse spectra show soft X-ray lines complexes which
indicate that most of the diffuse emission is due to hot diffuse
gas.

Perhaps the most intriguing new result from this observation is
the spiral feature in the diffuse emission near the center of the
galaxy.
We consider several models for the origin of the spiral feature including
absorption, thermal evaporation of cooler material, tidal effects,
cooling, adiabatic compression, and shocks.
Radiative cooling and shear might explain the feature, although
the spectral evidence is not consistent with this.
It seems most likely that this feature is due to adiabatic or
shock compression of ambient gas.
We suggest that the spiral is gas which is compressed at the edges
of two radio lobes.
This is consistent with the radio axis being perpendicular to the
dust lane in the nucleus of the galaxy.
Unfortunately, there is no high resolution radio image of this southern
galaxy;
thus, we are in the unusual situation (previous to $Chandra$, at least)
of having much higher resolution X-ray than radio data.
The radio flux is low enough that energetic arguments suggest
that the spiral feature can be explained by an interaction with the radio
source only if the present-day source is a fading remnant of previously
more intense radio activity. 

\acknowledgements
We are very grateful to Arunav Kundu for providing us with his
unpublished list of globular clusters in NGC~1553, for determining the
optical position of the nucleus in NGC~1553, and for many very
helpful conversations about the HST observations of this galaxy.
We thank Maxim Markevitch for several extremely helpful communications
concerning the background in the ACIS detector.
We are very grateful to Alexey Vikhlinin for providing his software
package for extracting X-ray spectra and constructing the response files
for extended sources.
We thank Shri Kulkarni for the interesting suggestion that all or
most LMXBs are formed in globular clusters.
Support for this work was provided by the National Aeronautics and Space
Administration through $Chandra$ Award Numbers
GO0-1019X,
GO0-1141X,
and
GO0-1173X,
issued by the $Chandra$ X-ray Observatory Center, which is operated by the
Smithsonian Astrophysical Observatory for and on behalf of NASA under
contract
NAS8-39073.
J. A. I. was supported by {\it Chandra} Fellowship grant PF9-10009, awarded
through the {\it Chandra} Science Center.

\begin{deluxetable}{rcccrrrcc}
\tablewidth{0pt}
\tablecaption{NGC~1553: Discrete X-ray Sources \label{tab:sources}}
\tablehead{
\colhead{Src.} & \colhead{Src.} &\colhead{RA(J2000)} & \colhead{Dec(J2000)} & \colhead{$d$} &
\colhead{Count Rate} & \colhead{SNR} & \colhead{$L_X$ (0.3-10 keV)} & \colhead{Notes}\\
\colhead{No.} & \colhead{Name} &\colhead{(h:m:s)} & \colhead{(\arcdeg:\arcmin:\arcsec)} &
\colhead{(\arcsec)} &
\colhead{($10^{-4}$ s$^{-1}$)} & \colhead{} &\colhead{($10^{38}$ erg s$^{-1}$)} & \colhead{}\\
\colhead{(1)} & \colhead{(2)} & \colhead{(3)} & \colhead{(4)} &
\colhead{(5)} & \colhead{(6)} & \colhead{(7)} &\colhead{(8)} & \colhead{(9)}}
\startdata
  1& CXOU J041610.5-554646 &  4:16:10.59 &    -55:46:46.8  &   0.0&  356.41$\pm12.50$& 28.5 &    175.07  \\
  2& CXOU J041609.9-554646 &  4:16:09.98 &    -55:46:46.6  &   5.2&   17.95$\pm2.90$&   6.2 &      8.82 \\
  3& CXOU J041610.4-554639 &  4:16:10.43 &    -55:46:39.0  &   7.9&   15.07$\pm2.67$&   5.6 &      7.40 \\
  4& CXOU J041609.6-554642 &  4:16:09.69 &    -55:46:42.2  &   8.9&    7.64$\pm1.94$&   3.9 &      3.75 & HST,var\\
  5& CXOU J041610.1-554637 &  4:16:10.14 &    -55:46:37.5  &  10.1&    9.04$\pm2.08$&   4.3 &      4.44 \\
  6& CXOU J041611.7-554652 &  4:16:11.76 &    -55:46:52.5  &  11.4&   10.02$\pm2.21$&   4.5 &      4.92 \\
  7& CXOU J041612.0-554658 &  4:16:12.05 &    -55:46:58.5  &  16.9&    5.47$\pm1.62$&   3.4 &      2.69 \\
  8& CXOU J041612.8-554647 &  4:16:12.86 &    -55:46:47.4  &  19.2&    8.46$\pm2.03$&   4.2 &      4.15 \\
  9& CXOU J041607.7-554643 &  4:16:07.75 &    -55:46:43.0  &  24.2&    4.79$\pm1.56$&   3.1 &      2.35 \\
 10& CXOU J041610.9-554621 &  4:16:10.92 &    -55:46:21.6  &  25.3&    4.42$\pm1.50$&   2.9 &      2.17 \\
 11& CXOU J041611.0-554712 &  4:16:11.08 &    -55:47:12.0  &  25.5&    4.56$\pm1.50$&   3.0 &      2.24 & HST\\
 12& CXOU J041609.5-554621 &  4:16:09.52 &    -55:46:21.1  &  27.3&    5.43$\pm1.62$&   3.3 &      2.67 \\
 13& CXOU J041606.1-554626 &  4:16:06.18 &    -55:46:26.1  &  42.6&    3.97$\pm1.37$&   2.9 &      1.95 \\
 14& CXOU J041608.3-554604 &  4:16:08.31 &    -55:46:04.9  &  46.2&    4.28$\pm1.44$&   3.0 &      2.10 \\
 15& CXOU J041607.1-554610 &  4:16:07.15 &    -55:46:10.1  &  46.8&   11.03$\pm2.26$&   4.9 &      5.42 \\
 16& CXOU J041616.3-554618 &  4:16:16.38 &    -55:46:18.4  &  56.5&   40.04$\pm4.22$&   9.5 &     19.67 \\
 17& CXOU J041602.9-554655 &  4:16:02.90 &    -55:46:55.1  &  65.4&   66.51$\pm5.41$&  12.3 &     32.67 & var\\
 18& CXOU J041611.7-554753 &  4:16:11.77 &    -55:47:53.5  &  67.5&    7.24$\pm1.83$&   4.0 &      3.56 \\
 19& CXOU J041615.6-554741 &  4:16:15.65 &    -55:47:41.5  &  69.4&    4.46$\pm1.52$&   2.9 &      2.19 \\
 20& CXOU J041612.2-554516 &  4:16:12.25 &    -55:45:16.7  &  91.1&    4.94$\pm1.52$&   3.2 &      2.43 \\
 21& CXOU J041608.3-554816 &  4:16:08.31 &    -55:48:16.8  &  92.0&   29.11$\pm3.58$&   8.1 &     14.30 \\
 22& CXOU J041559.2-554615 &  4:15:59.26 &    -55:46:15.2  & 100.7&   10.74$\pm2.24$&   4.8 &      5.27 & DSS\\
 23& CXOU J041621.5-554731 &  4:16:21.52 &    -55:47:31.9  & 102.7&    6.14$\pm1.68$&   3.7 &      3.02 & HST\\
 24& CXOU J041611.5-554832 &  4:16:11.51 &    -55:48:32.9  & 106.4&    9.79$\pm2.12$&   4.6 &      4.81 \\
 25& CXOU J041623.2-554725 &  4:16:23.21 &    -55:47:25.3  & 113.2&    6.56$\pm1.74$&   3.8 &      3.22 \\
 26& CXOU J041620.9-554805 &  4:16:20.99 &    -55:48:05.3  & 117.7&   12.56$\pm2.37$&   5.3 &      6.17 \\
 27& CXOU J041604.1-554458 &  4:16:04.17 &    -55:44:58.6  & 121.0&    9.46$\pm2.22$&   4.3 &      4.65 & DSS\\
 28& CXOU J041602.0-554508 &  4:16:02.00 &    -55:45:08.1  & 122.5&    5.85$\pm1.72$&   3.4 &      2.88 \\
 29& CXOU J041616.5-554839 &  4:16:16.51 &    -55:48:39.3  & 123.0&   17.25$\pm2.77$&   6.2 &      8.47 \\
 30& CXOU J041615.6-554425 &  4:16:15.67 &    -55:44:25.4  & 147.8&   11.92$\pm2.40$&   5.0 &      5.86 \\
 31& CXOU J041630.5-554613 &  4:16:30.52 &    -55:46:13.6  & 171.4&   20.32$\pm3.03$&   6.7 &      9.98 \\
 32& CXOU J041631.3-554703 &  4:16:31.31 &    -55:47:03.8  & 175.6&    8.89$\pm2.02$&   4.4 &      4.37 \\
 33& CXOU J041602.6-554932 &  4:16:02.65 &    -55:49:32.4  & 178.6&   11.34$\pm2.26$&   5.0 &      5.57 \\
 34& CXOU J041615.9-554949 &  4:16:15.94 &    -55:49:49.6  & 188.3&    7.86$\pm1.89$&   4.2 &      3.86 & DSS\\
 35& CXOU J041602.8-554947 &  4:16:02.83 &    -55:49:47.8  & 192.4&   23.96$\pm3.40$&   7.0 &     11.77 & DSS\\
 36& CXOU J041617.9-554955 &  4:16:17.91 &    -55:49:55.1  & 198.2&    7.97$\pm1.90$&   4.2 &      3.91 & var\\
 37& CXOU J041626.0-554406 &  4:16:26.05 &    -55:44:06.5  & 206.7&    5.38$\pm1.69$&   3.2 &      2.64 \\
 38& CXOU J041551.5-554859 &  4:15:51.53 &    -55:48:59.7  & 208.5&  193.17$\pm9.71$&  19.9 &     94.89 & DSS\\
 39& CXOU J041547.3-554810 &  4:15:47.38 &    -55:48:10.3  & 212.8&    5.94$\pm1.72$&   3.5 &      2.92 \\
 40& CXOU J041545.5-554543 &  4:15:45.56 &    -55:45:43.9  & 220.4&   12.48$\pm2.55$&   4.9 &      6.13 & var\\
 41& CXOU J041611.5-555029 &  4:16:11.50 &    -55:50:29.4  & 222.8&   12.16$\pm2.36$&   5.2 &      5.97 & var\\
 42& CXOU J041635.5-554440 &  4:16:35.56 &    -55:44:40.5  & 245.7&    8.55$\pm2.08$&   4.1 &      4.20 \\
 43& CXOU J041637.9-554520 &  4:16:37.90 &    -55:45:20.7  & 246.0&    6.64$\pm1.87$&   3.6 &      3.26 \\
 44& CXOU J041640.4-554510 &  4:16:40.41 &    -55:45:10.5  & 269.5&    8.30$\pm2.04$&   4.1 &      4.08 \\
 45& CXOU J041538.2-554539 &  4:15:38.25 &    -55:45:39.1  & 281.1&    5.98$\pm2.05$&   2.9 &      2.94 & DSS\\
 46& CXOU J041540.3-554847 &  4:15:40.37 &    -55:48:47.0  & 281.7&    7.84$\pm2.27$&   3.5 &      3.85 \\
 47& CXOU J041643.2-554510 &  4:16:43.21 &    -55:45:10.5  & 291.6&    9.84$\pm2.25$&   4.4 &      4.83 \\
 48& CXOU J041603.3-554110 &  4:16:03.31 &    -55:41:10.7  & 341.7&   43.52$\pm4.90$&   8.9 &     21.38 & DSS\\
 49& CXOU J041557.2-554123 &  4:15:57.21 &    -55:41:23.4  & 342.5&    8.18$\pm2.32$&   3.5 &      4.02 \\
\enddata								    
\tablecomments{In column (9), ``HST'' indicates that the source is identified
with a globular cluster from HST data,
``DSS'' indicates that there is an apparent optical identification on the
DSS image,
and ``var'' means that the source may be variable.}
\end{deluxetable}							    

\clearpage

\begin{deluxetable}{lccccccc}
\tablenum{2}
\tablewidth{0pt}
\tablecaption{Spectral Fits\label{tab:spectra}}
\tablehead{
\colhead{Component} & \colhead{$N_H$} & \colhead{Soft Model} & \colhead{$kT_s$} & 
\colhead{Abund.} &  \colhead{$\Gamma$} & 
\colhead{$\chi^{2}$/d.o.f.} & \colhead{Total Cts.}\\
\colhead{} & \colhead{($10^{20}$ cm$^{-2}$)} & \colhead{} & \colhead{(keV)} & 
\colhead{(solar)} & \colhead{} & 
\colhead{} & \colhead{(0.7-10.0 keV)}}
\startdata

total &  (1.41) & mekal & 0.48$^{+0.05}_{-0.10}$ &
0.05$^{+0.06}_{-0.03}$ & 0.63$^{+0.33}_{-0.19}$ & 315.6/260& 3616\\

total & (1.41) & mekal & 0.47$^{+0.04}_{-0.13}$ &
0.16$^{+0.33}_{-0.05}$ & (1.20) & 321.9/261 & 3616\\

diffuse & (1.41) & mekal & 0.48$^{+0.04}_{-0.09}$ &
0.06$^{+0.06}_{-0.04}$  &
-0.35$^{+0.86}_{-1.32}$  & 263.4/229 & 2328\\

diffuse & (1.41) & mekal & 0.47$^{+0.04}_{-0.10}$ &
0.10$^{+0.06}_{-0.05}$  &
(1.20) & 270.8/230 & 2328\\

sources & (1.41) & bbody & 0.24$^{+0.05}_{-0.07}$ &&
1.20$^{+0.39}_{-0.15}$ & 
35.2/40 & 1332 \\

AGN & 18.25$^{+7.42}_{-7.59}$ & diskbb & 1.70$^{+0.39}_{-0.22}$ &
& &34.7/32 & 758\\

AGN & 36.04$^{+12.95}_{-9.22}$ & & & & 1.59$^{+0.11}_{-0.18}$ & 40.3/32 &758\\

spiral & (1.41) & mekal & 0.51$^{+0.07}_{-0.09}$ & 
0.12$^{+0.17}_{-0.07}$ & 
-1.55$^{+1.57}_{-0.25}$ & 19.5/24 & 465\\

spiral & (1.41) & mekal & 0.50$^{+0.06}_{-0.06}$ & 
0.38$^{+0.28}_{-0.14}$ & 
(1.20) & 27.1/25 & 465 \\

central circle & (1.41) & mekal & 0.48$^{+0.07}_{-0.09}$ &
0.10$^{+0.34}_{-0.07}$ & 
1.38$^{+1.54}_{-2.12}$ & 38.6/43 & 572\\

central circle & (1.41) & mekal & 0.49$^{+0.07}_{-0.08}$ &
0.10$^{+0.06}_{-0.04}$ & 
(1.20) & 38.7/44 & 572\\

\enddata                                                                    
\end{deluxetable}

\clearpage

\clearpage

\clearpage

\clearpage
\begin{figure}[t!]
\plotone{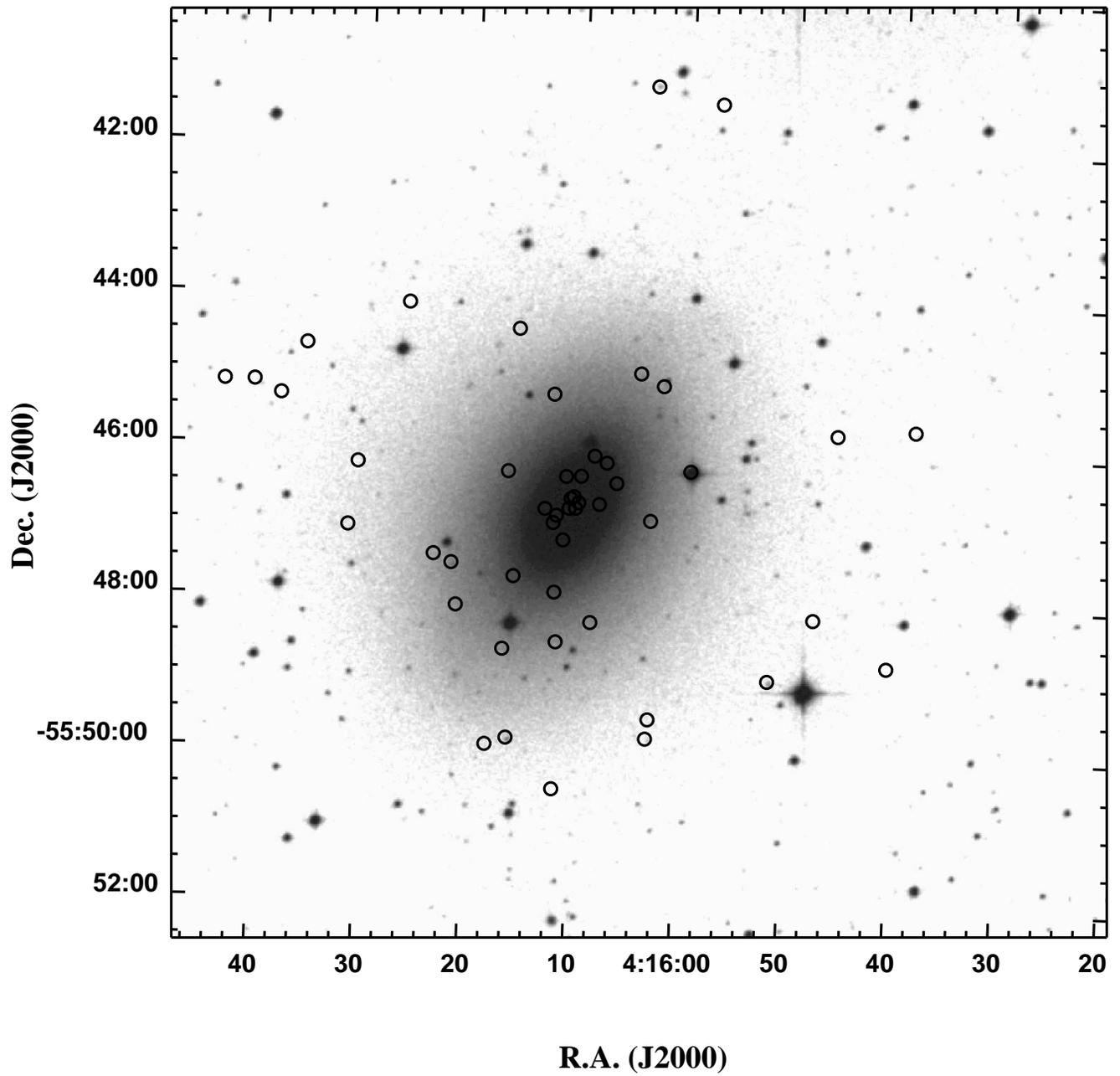}
\caption{The positions of the X-ray sources are
overlaid as circles onto the DSS optical image of NGC~1553.
\label{fig:optical}}
\end{figure}

\clearpage

\clearpage

\clearpage

\clearpage

\clearpage

\clearpage

\end{document}